\documentclass[twocolumn,showpacs,preprintnumbers,amsmath,amssymb]{revtex4}

\usepackage{graphicx}
\usepackage{dcolumn}
\usepackage{bm}
\usepackage{enumerate}
\usepackage[latin1]{inputenc}
\newcommand{\comment}[1]{}
\newcommand\etal{\mbox{\textit{et al.~}}}
\usepackage{graphicx,colordvi}
\usepackage[usenames]{color}
\usepackage{bm}

\newcommand{\refeq}[1]{(\ref{#1})}

\begin{document}
\setlength{\unitlength}{0.7\textwidth}
\title{The decay of magnetohydrodynamic turbulence in a confined domain}
\author{Salah Neffaa$^{1}$, Wouter J.T.  Bos$^{1,2}$, Kai Schneider$^{1}$} 
\affiliation{$^1$ M2P2-CNRS and CMI, Universit\'e de Provence, 38 rue Fr\'ed\'eric Joliot-Curie, 13451 Marseille cedex 20, France\\
$^2$ LMFA, UMR CNRS 5509, Ecole Centrale de Lyon - Universit\'e Claude
Bernard Lyon 1 - INSA de Lyon, 69134 Ecully cedex, France}

\date{\today}
%
\begin{abstract}
The effect of non periodic boundary conditions on decaying two-dimensional magnetohydrodynamic turbulence is investigated. We consider a circular domain with no-slip boundary conditions for the velocity and where the normal component of the magnetic field vanishes at the wall. Different flow regimes are obtained by starting from random initial velocity and magnetic fields with varying integral quantities. These regimes, equivalent to the ones observed by Ting, Matthaeus and Montgomery [Phys. Fluids {\bf 29}, 3261, (1986)] in periodic domains, are found to subsist in confined domains. We examine the effect of solid boundaries on the energy decay and alignment properties. The final states are characterized by functional relationships between velocity and magnetic field.

\end{abstract}

\maketitle
\section{Introduction}

The influence of initial conditions on decaying magnetohydrodynamic (MHD) turbulence received considerable interest in the 1980's, because of its relevance to explain solar-wind data \cite{Dobrowolny1980,Grappin1982,Matthaeus1983}. Indeed, in magnetohydrodynamics the behavior of decaying turbulent flow depends strongly on the initial conditions, and different initial values and ratios of integral quantities can lead to a wide variety of distinct behaviors. The first systematic study of the different possible types of decay was performed by Ting, Matthaeus and Montgomery \cite{Ting1986}, who identified four classes of possible decay behavior, corresponding roughly to a magnetically dominated, a hydrodynamically dominated, a magnetically-hydrodynamically equipartitioned and an erratic transition regime. Their study considered the two-dimensional case, which is not only relevant in applications in which an externally imposed field renders the flows quasi two-dimensional, but also from a general physical understanding of MHD turbulence, which behaves quite similar in two and three dimensions, due to the equivalent role of the ideal invariants \cite{Biskamp2001}.

Whereas the influence of the initial conditions on decaying MHD turbulence has been studied and understood to some extend, studies on the effect of boundary conditions have been limited to low resolutions \cite{Shan1994,Mininni2006,Mininni2007}, imposed by the numerical methods used to account for boundaries. Even though these investigations highlighted interesting physics, higher resolution simulations are needed to obtain a better understanding of wall-bounded MHD, which plays a dominant role  in geophysical flows in the core of planets such as the earth and industrial processes involving liquid metals. For the hydrodynamic case it was found that, boundary conditions have a significant influence on two-dimensional turbulence \cite{Schneider2005-2}. In contrast to the periodic domain, where generally a long lasting  state is found with a functional $sinh$ relationship between the vorticity and the stream function \cite{Joyce1973}, corresponding to two counterrotating vortices, in a bounded domain with no-slip wall conditions the final state yields an axisymmetric vorticity distribution, for which a linear relation between vorticity and stream function is observed \cite{Schneider2007}.

 \vspace{1.5cm}

\begin{figure}[h]
\unitlength=0.02cm
\begin{picture}(200,100)
\put(-150,0){
\put(200,1){ \includegraphics[height=150\unitlength]{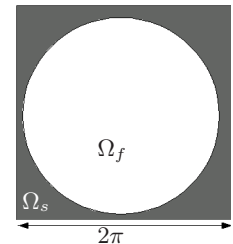}}
\put(260,-8){$2\pi$}
\put(260,50){$\Omega_f$}
\put(210,15){{\color{white} $\Omega_s$}}}
\end{picture}
 \caption{The computational domain is a square box $2\pi$. The fluid domain $\Omega_f$ is a circular container with radius $r=\frac{19}{20}\pi$, surrounded by the solid domain $\Omega_s$.\label{domain}}
\end{figure}

 In the present work we propose an extension of the volume penalization method \cite{Angot1999} to two-dimensional MHD to compute decaying flows in bounded domains using an efficient Fourier pseudo-spectral method. We address the following questions: what is the influence of confinement by fixed solid boundaries on decaying two-dimensional MHD turbulence? Do the four regimes found by Ting \etal \cite{Ting1986} continue to exist in the presence of boundaries? What are the final (viscously decaying) states?

\section{Governing equations and numerical method}

We consider resistive MHD, formulated in usual dimensionless variables ${\bf{u}}=(u,v)$ and ${\bf{B}}=(B_x,B_y)$ which are respectively the velocity and the magnetic field. The flow is considered to be two-dimensional, incompressible and we assume the mass density to be constant.\\The governing equations are the following:
\begin{eqnarray} 
\label{qte_mvt}
\frac{\partial{\bf{u}}}{\partial{t}}+\bf{u}\cdot\nabla\bf{u}= -\nabla p + \bf{j}\times\bf{B} + \nu\nabla^2\bf{u} -\frac{1}{\epsilon}\chi(\bf{u} - \bf{u_0})\\
\label{eq_magnetic}
 \frac{\partial{\bf{B}}}{\partial{t}} = \nabla\times(\bf{u}\times\bf{B}) + \eta\nabla^2\bf{B}-\frac{1}{\epsilon}\chi(\bf{B} - \bf{B_0})\\
\label{DivFree}
\nabla\cdot\textbf{u} = 0 \hspace{1cm} \nabla\cdot\textbf{B} = 0
\end{eqnarray}

Here $\nu$ and $\eta$ are respectively the kinematic viscosity and the magnetic diffusivity. $\omega\bf{e}_z=\nabla\times\bf{u}$ is the vorticity, $j\bf{e}_z=\nabla\times\bf{B}$ is the current density. Furthermore we define the vector potential $\bf{a} = \textrm{a}\bf{e}_z$ as $\bf{B} = \nabla\times\bf{a}$ and the stream function $\psi$ as ${\bf u} = \nabla^{\perp}\psi=(-\partial{\psi}/\partial{y},\partial{\psi}/\partial{x})$. An originality in our approach is the way in which the boundary conditions are imposed: we use volume (or surface in 2D) penalization \cite{Angot1999,Schneider2005-3} to include the boundary conditions.  This method has the advantage that arbitrary basis-functions can be used. In our case a Fourier pseudo-spectral code is employed. The advantage with respect to a method based on a decompostion in terms of Chandrasekhar-Kendall eigenfunctions \cite{Shan1994,Mininni2006,Mininni2007} is, that fast Fourier transforms can be used, allowing for high resolution computations of low computational cost. Also, its application to three dimensional flows is conceptually straightforward and will be addressed in a future work. The additional terms on the right hand side of equation (\ref{qte_mvt}) and (\ref{eq_magnetic}) correspond to this penalization-method. The quantities $\bf{u_0}$ and $\bf{B_0}$ correspond to the values imposed in the solid part of the numerical domain $\Omega_s$, illustrated in figure \ref{domain}. Here we choose $\bf{u_0}=\bf{0}$ and $\bf{B_0}={\bf B_\parallel}$ (where ${\bf B_\parallel}$ is the tangential component of ${\bf B}$ at the wall), corresponding to vanishing velocity and no penetration of magnetic field into the solid domain which is hence considered as a perfect conductor, coated inside with a thin layer of insulant, which guarantees that the current density cannot penetrate into the solid \cite{Mininni2006}. The mask function $\chi$ is equal to 0 inside $\Omega_f$ (where the penalization terms thereby dissappear) and equal to 1 inside $\Omega_s$.  The physical idea is to model the solid part as a porous medium whose permeability $\epsilon$ tends to zero \cite{Angot1999,Schneider2005-3}. For $\epsilon$ $\rightarrow$ 0, where the obstacle is present, the velocity $\bf{u}$ tends to $\bf{u}_0$ and the magnetic field $\bf{B}$ tends to $\bf{B}_0$. The nature of the boundary condition for the velocity is thus no-slip at the wall.

In the two-dimensional case it is convenient to take the curl of \refeq{qte_mvt} and \refeq{eq_magnetic} to obtain after simplification equations for the vorticity and current density. These are scalar equations which automatically satisfy the incompressibility conditions \refeq{DivFree}. The equations are then
\begin{eqnarray} \label{eq_vorticite}
\noindent \frac{\partial{\omega}}{\partial{t}} = -{\bf u}\cdot\nabla\omega + {\bf B}\cdot\nabla j + \nu\nabla^2\omega -\frac{1}{\epsilon}\nabla\times[\chi(\bf{u}-\bf{u_0})]\\
\label{eq_induction}
 \frac{\partial{j}}{\partial{t}} = -\nabla^2 \left(\textbf{u}\times\bf{B}\right) + \eta\nabla^2 j  - \frac{1}{\epsilon}\nabla\times[\chi(\bf{B} - \bf{B_0})]~~~~~~
\end{eqnarray}

The equations are discretized with a classical Fourier pseudo-spectral method imposing periodic boundary conditions on the square domain of size $2\pi$, using $512^2$ grid points. At each iteration the fields are dealiased by spherical truncation following the $2/3$ rule. The penalization parameter $\epsilon$, corresponding to the permeability of the solid domain, is taken equal to $10^{-3}$, a value validated by a systematic study of the sensitivity of the results to this parameter \cite{Schneider2005-3}. The fluid viscosity $\nu$ and magnetic diffusivity $\eta$ were taken equal to $10^{-3}$, the timestep $dt$ equals to $5.10^{-4}$. The initial kinetic and magnetic Reynolds number are defined as $Re=2r\sqrt{2E_u(t=0)}/\nu$ and $Re_m=2r\sqrt{2E_B(t=0)}/\eta$, where $r$ is the radius of the domain and, $E_u$ and $E_B$ are the kinetic and magnetic energies, respectively (see table \ref{init_val}).

\section{Initial conditions}

Both vorticity and current density fields are initialized with Gaussian random initial conditions. Therefore, their Fourier transforms $\widehat{\omega}$ and $\widehat{j}$, where $\widehat{\omega}(\bf{k}) = \frac{1}{4\pi^2}\int\omega(\bf{x}) e^{-\imath{\bf{k}}\cdot{\bf{x}}}dx$, are initialized with random phases and, their amplitudes give the energy spectra:
\begin{equation}
 E_u(k), E_B(k) \propto \frac{k}{(g + (k/k_0))^4}
\end{equation}
with $k=\vert{\bf k}\vert$ and, where $g=0.98$ and $k_0=\frac{3}{4}\sqrt{2}\pi$. This energy spectrum follows a power law proportional to $k^{-3}$ at large wavenumbers and was chosen to compare with simulations performed in the periodic case. Both fields are statistically identical. The corresponding fields $\bf{u}$ and $\bf{B}$ are calculated from $\omega$ and $j$ using the Biot-Savart law.

For vanishing viscosity and resistivity, two-dimensional MHD has three conserved invariants. The total energy is $E$, defined as sum of the kinetic energy $E_u$ and the magnetic energy $E_B$ :
\begin{equation} \label{NRJ}
 E = E_u + E_B = \frac{1}{2}\int_{\Omega_f}(\vert\textbf{u}\vert^2 + \vert\textbf{B}\vert^2)~d^2x 
\end{equation}
$H_c$ is the cross helicity:
\begin{equation} \label{HELIC}
 H_c = \frac{1}{2}\int_{\Omega_f} \textbf{u}\cdot\textbf{B}~d^2x
\end{equation}
which mesures the global correlations between $\bf{u}$ and $\bf{B}$ and $A$ is the integral of the squared vector potential:
\begin{equation} \label{POT}
 A = \frac{1}{2}\int_{\Omega_f} a^2~d^2x
\end{equation}
As was shown by Ting \etal \cite{Ting1986} for periodic boundary conditions, the dynamics of decaying MHD turbulence depend strongly on the initial values of these invariants. Because of its interest for the present study we recall briefly the four distinct decay regimes discerned by Ting \etal \cite{Ting1986} depending on the initial values and ratios of the invariants. First, in the case of small initial $H_c$ and $E_B>E_u$, a magnetically dominated regime is obtained. 
Selective decay is observed in this regime which corresponds to the decay of $E_u$ relative to $E_B$. Second, in the case of vanishingly small initial magnetic energy, the Lorentz force acting in the vorticity equation can not become strong enough, so that the vector potential is advected like a passive scalar. Following Biskamp and Welter \cite{Biskamp1990}, the magnetic field may however be amplified even if the initial ratio $E_B/E_u$ is very small, given that $\eta$ is sufficiently small. They found that $E_u/E_B < Re_m^2$ is necessary such that the magnetic field can be intensified. This is a regime which essentially corresponds to the Navier-Stokes limit. Third, in the case of substantial initial cross-helicity, the turbulence tends towards an Alfv\'enic state in which ${\bf u}$ and ${\bf B}$ are aligned or anti-aligned and approximately equipartitioned. This process is called dynamic alignment and the ratio $E/\vert H_c\vert$ tends to two. This is a state free from nonlinear interactions, inhibiting cascade processes (even though this depletion of nonlinearity is rather slow with increasing cross-helicity \cite{Matthaeus1983}). The fourth and final regime is an erratic regime which might tend to different final states and which could be related to various competing subregions with unequal sign of cross-helicity. This regime can be found if the flow is initialized with small cross-helicity and comparable kinetic and magnetic energies.

Whether these regimes persist in the presence of solid boundaries is one of the main questions we want to answer in the present work. To obtain the desired initial conditions corresponding to the four regimes we proceed as follows:

Starting from random initial conditions in Fourier space, we renormalize ${\bf u}$ and ${\bf B}$ in physical space  by varying the coefficient $\alpha$: 
\begin{equation} \label{ReNormal}
 \bf{u}^* = \frac{\alpha}{\sqrt{2E_u}}\bf{u} \hspace{2cm} \bf{B}^* = \frac{1}{\sqrt{2E_B}}\bf{B}
\end{equation}
This generally yields initial conditions with vanishingly small cross-helicity, and initial conditions for regime I,II and IV can hereby be created. In the case of regime III, a non-zero cross-helicity needs to be imposed. We achieve this by creating a random initial condition for ${\bf u}$ and a perpendicular field ${\bf {\bf u}_{\perp}}$ by rotating ${\bf u}$ by $\pi/2$. The magnetic field is then obtained by a linear combination of the 2 fields:
\begin{equation} \label{Rotate}
 \bf{B}^* = \beta{\bf{u}} + (1-\beta){\bf{u}_{\perp}}
\end{equation}
Hereby any given cross helicity can be imposed. Table \ref{init_val} summarizes the initial values of $E/A$, $E_u/E_B$ and $H_c$ for the four different regimes, together with the Reynolds number.

\begin{table}
\begin{center}
\begin{tabular}{l||c|c|c|c|c}
 \hline 
 &  ~~~~E/A~~ & $E_u/E_B$ & $H_c$ & ~$Re$~ & ~$Re_m$~\\ \hline
Regime I   & 14.4 & 0.068 & 0.05 & 3580 & 7310\\ 
Regime II  & $3\cdot10^{5}$ & $2.6\cdot10^{4}$  & $3.6\cdot10^{-5}$ & 7900 & 53\\ 
Regime III & 53.5 & 1.48  & 0.23 & 5340 & 4620\\ 
Regime IV  & 14.03 & 1.18  & $9.5\cdot10^{-3}$ & 5970 & 5340\\ 
\end{tabular}
\end{center}
\caption{\label{init_val}Initial values of the four different regimes.}
\end{table}

\section{Results and discussion}
\nointerlineskip
\subsection{Characterization of the different decay regimes}
\nointerlineskip
\begin{figure}
\setlength{\unitlength}{\textwidth}
\includegraphics[width=0.4\unitlength]{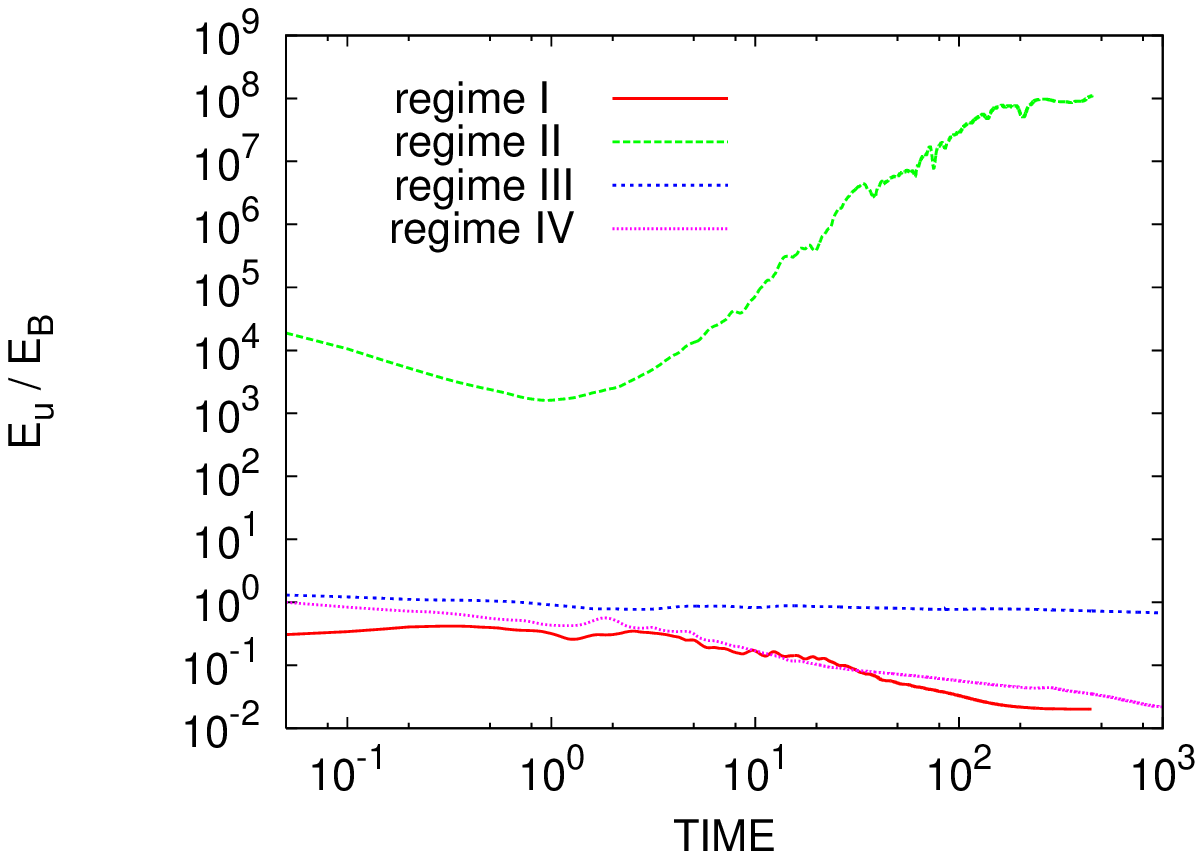}
\includegraphics[width=0.4\unitlength]{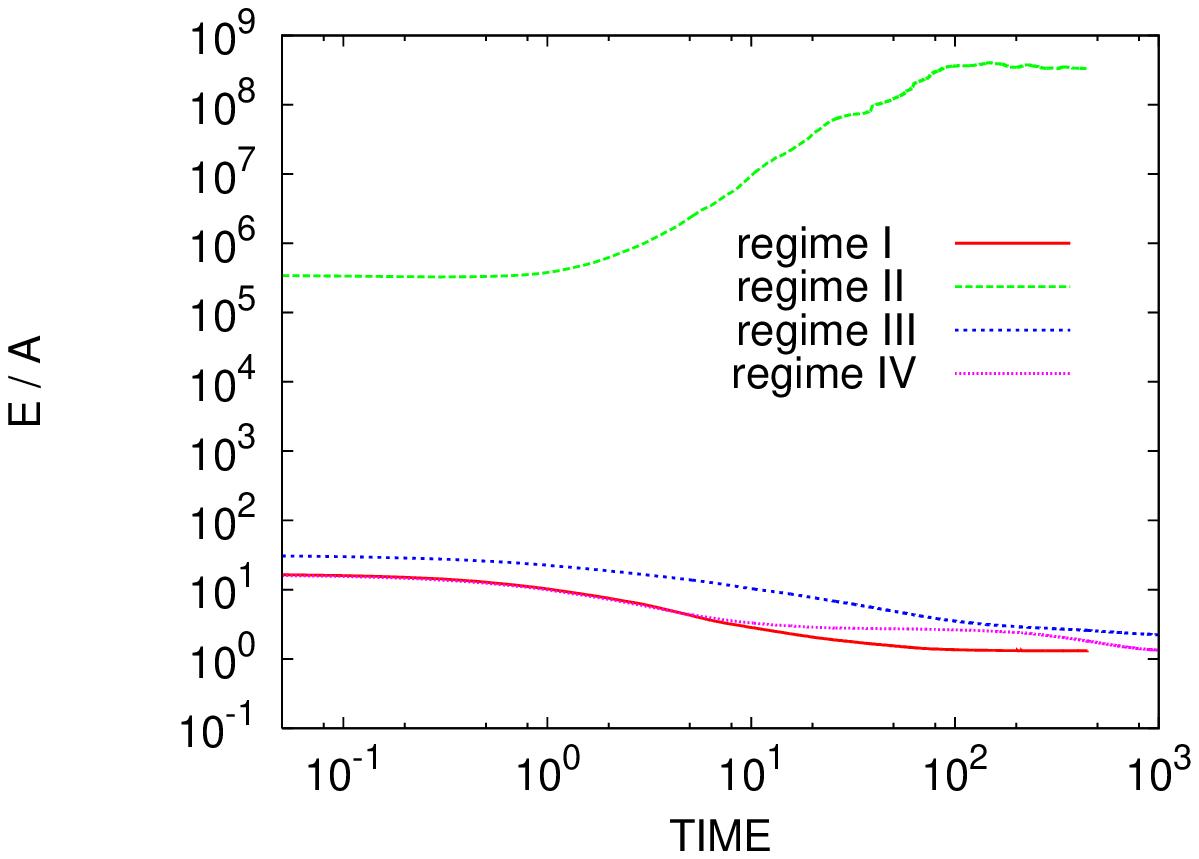}
\includegraphics[width=0.4\unitlength]{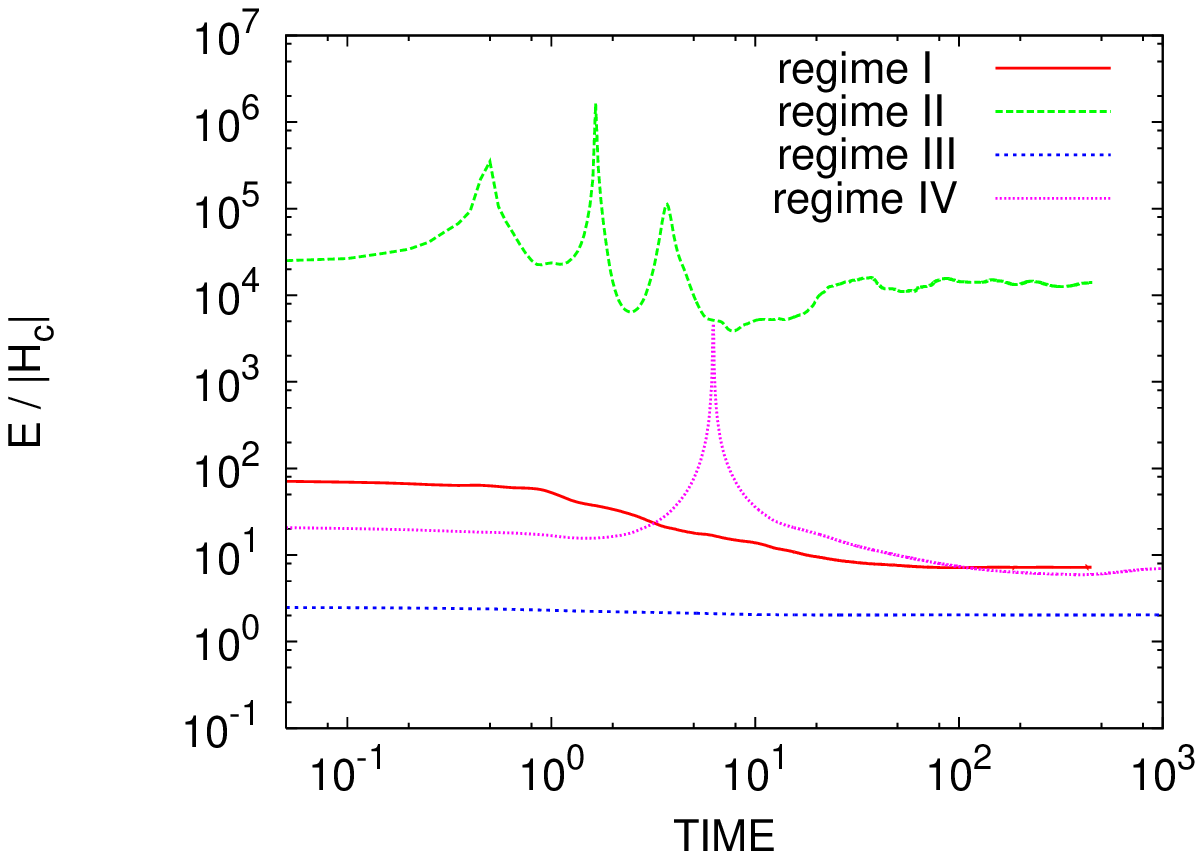}
\caption{Time evolution of integral quantities. Top: ratio of kinetic and magnetic energy, $E_u/E_B$. Center: ratio of total energy and integral of the squared vector potential, $E/A$. Bottom: ratio of total energy and magnetic helicity, $E/\vert H_c\vert$. \label{E_sur_A}}
\end{figure}

\begin{figure}
\setlength{\unitlength}{\textwidth}
 \includegraphics[width=0.4\unitlength]{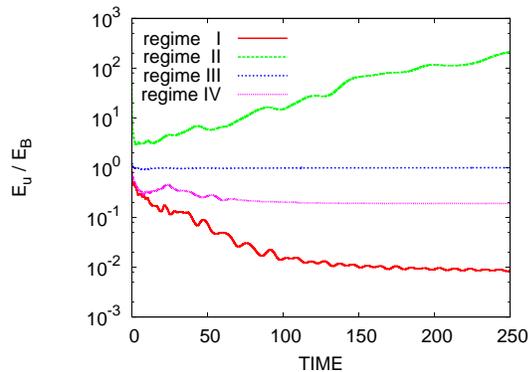}
\caption{Time evolution of $E_u/E_B$ in a periodic domain, starting from similar initial conditions as in figure \ref{E_sur_A}. \label{energy_per}}
\end{figure}
In figure \ref{E_sur_A} we show the time evolution of several integral quantities for the 4 different sets of initial conditions. The main observation is that the 4 different regimes, discerned by Ting \etal \cite{Ting1986} are robust enough to survive within a bounded domain. We now discuss the results in more detail.

 Regarding the ratio of kinetic and magnetic energy (Fig. \ref{E_sur_A}, top), it is observed that in the absence of initial cross-helicity (case I, II and IV) the magnetic energy finally dominates, unless it is very small initially (Navier Stokes limit). However, if the initial cross-helicity is initially large and $E_{u}/E_B$ is of order unity, the flow energy will remain approximately equipartitioned between the velocity and magnetic field. 

This picture is confirmed by the time evolution of the  ratio $E/A$ (Fig. \ref{E_sur_A}, center). In this representation it is however emphasized that in the Navier-Stokes limit (case II), the character of the magnetic field has changed: in the ideal system (vanishing viscosity and magnetic diffusivity), $A$ is a quantity that cascades towards the small wavenumbers. In a non-ideal system an inverse cascade generally slows down the dissipation rate of the quantity. However, in the limit of small Lorentz force, the equations of the vorticity and vector potential become equivalent to the equations that describe a passive scalar advected by a two-dimensional velocity field. The passive scalar being a quantity which cascades towards higher wavenumbers, the vector potential gets dissipated faster in this case than in the case where the Lorentz force is significant. This results in a rapid increase of the quantity $E/A$ in case II.

The ratio $E/\vert H_c\vert$ (Fig. \ref{E_sur_A}, bottom) attains its minimum absolute value $2$ for case III. This corresponds to dynamic alignment: the velocity field is equal in magnitude and perfectly aligned, or anti-aligned with the magnetic field. The erratic regime is clearly represented by case IV, in which the cross-helicity approaches a value close to zero. As we will see in the following, this is caused by different subregions with oppositely valued $H_c$.

For comparison, we show in figure \ref{energy_per} $E_{u}/E_B$ in a periodic domain, starting from similar initial conditions as in the bounded case using the same numerical parameters. Even though the trend is similar, we see that a more oscillatory behavior for case I and II is observed than in the case of the bounded domain. This oscillatory behavior is related to energy exchange between the magnetic field and the velocity field by means of Alfv\'en waves \cite{Pouquet1988}. Whereas in a periodic domain these waves can freely propagate, in a bounded domain they might be more rapidly suppressed, explaining the less oscillatory behavior of $E_u/E_B$ in a bounded domain. Further research is needed to clarify this.

The quantity $E/H_c$ gives a measure for the dynamic alignment, which corresponds to measuring both the equipartitioning of energy and the alignment properties. If we are exclusively interested in the alignment properties, the relative cross helicity, which corresponds to the cosine of the angle $\theta$ between the velocity and magnetic field vector,
\begin{equation}
 \cos\theta = \frac{H_c}{(E_u E_B)^{1/2}}
\end{equation}
should be considered.
\begin{figure}
\setlength{\unitlength}{\textwidth}
\includegraphics[width=0.4\unitlength]{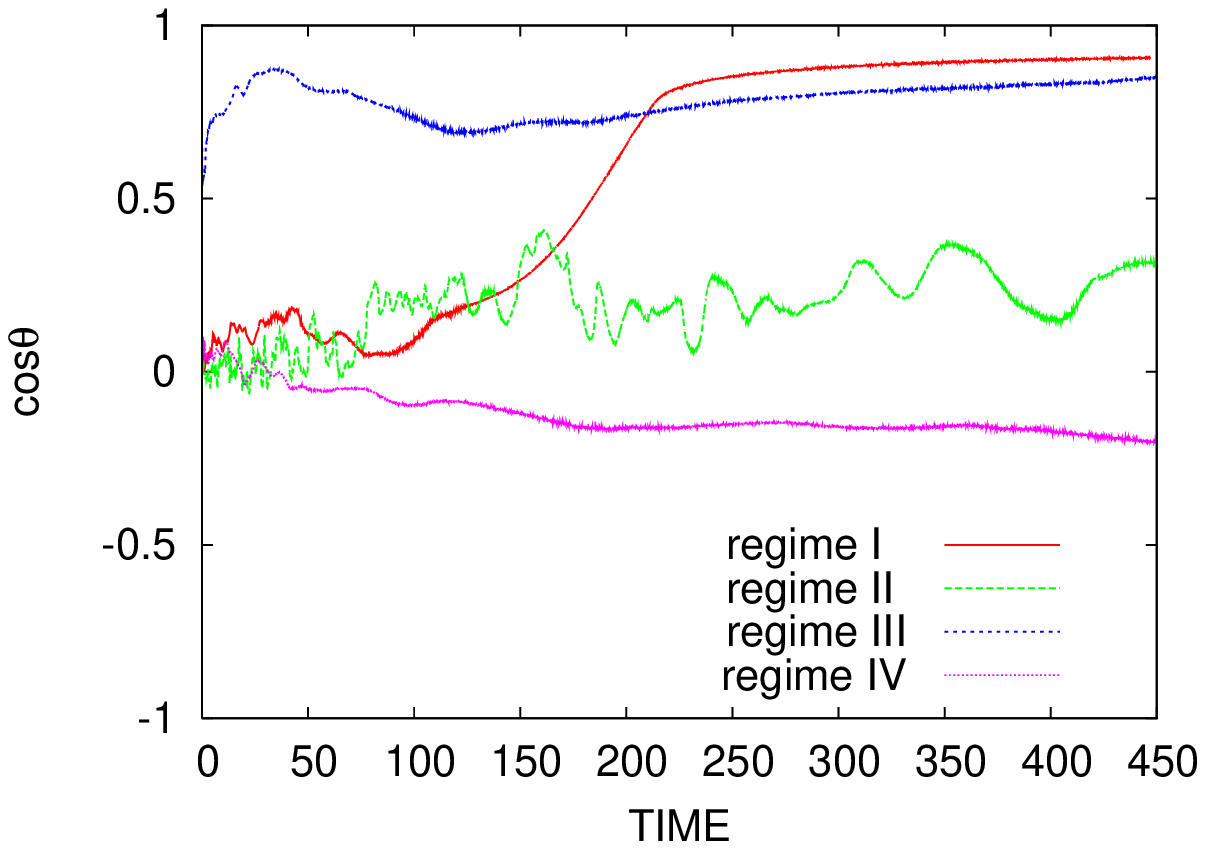}
\caption{Time evolution of the average alignment $\cos\theta$, between the magnetic field and the velocity field.\label{cost}}
\includegraphics[width=0.4\unitlength]{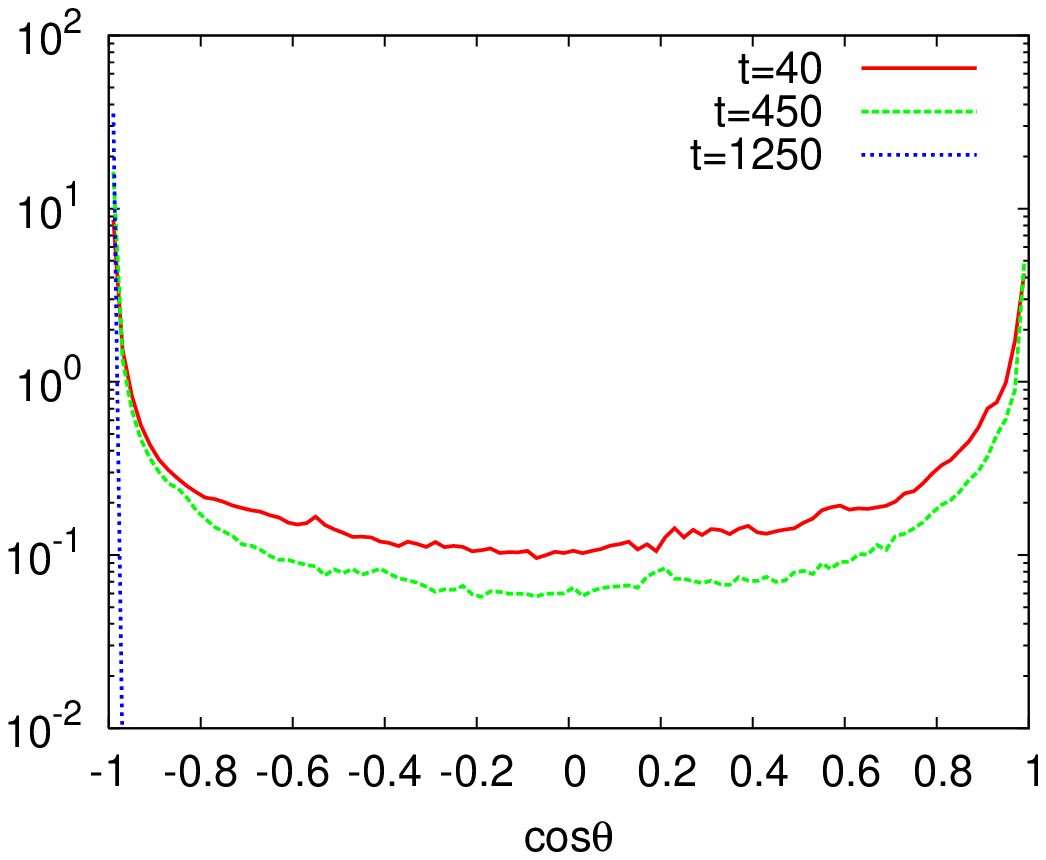}
\caption{Probability density of $\cos\theta$ at $t=40$, $t=450$ and $t=1250$ in the regime IV.\label{PDF_theta}}

\end{figure}

In figure \ref{cost}, $\cos\theta$ is plotted as a function of time. It can be observed that in case I and III, the velocity field tends to a nearly aligned state. In case II and IV, this quantity remains close to zero, however for a different reason. In case II, the alignment is small, because the vector potential is advected as a nearly passive scalar. In case IV the local alignment is large but different aligned or anti-aligned regions cancel out the contributions, yielding a net-global alignment close to $0$. This can be observed in the corresponding probability distribution function of $\cos\theta$ at $t=40$ and $t=450$, shown in figure \ref{PDF_theta}. Nevertheless, for long time ($t=1250$) we observe an anti-alignment.

\subsection{Energy decay and visualizations}
\nointerlineskip

\begin{figure}
\setlength{\unitlength}{\textwidth}
\includegraphics[width=0.4\unitlength]{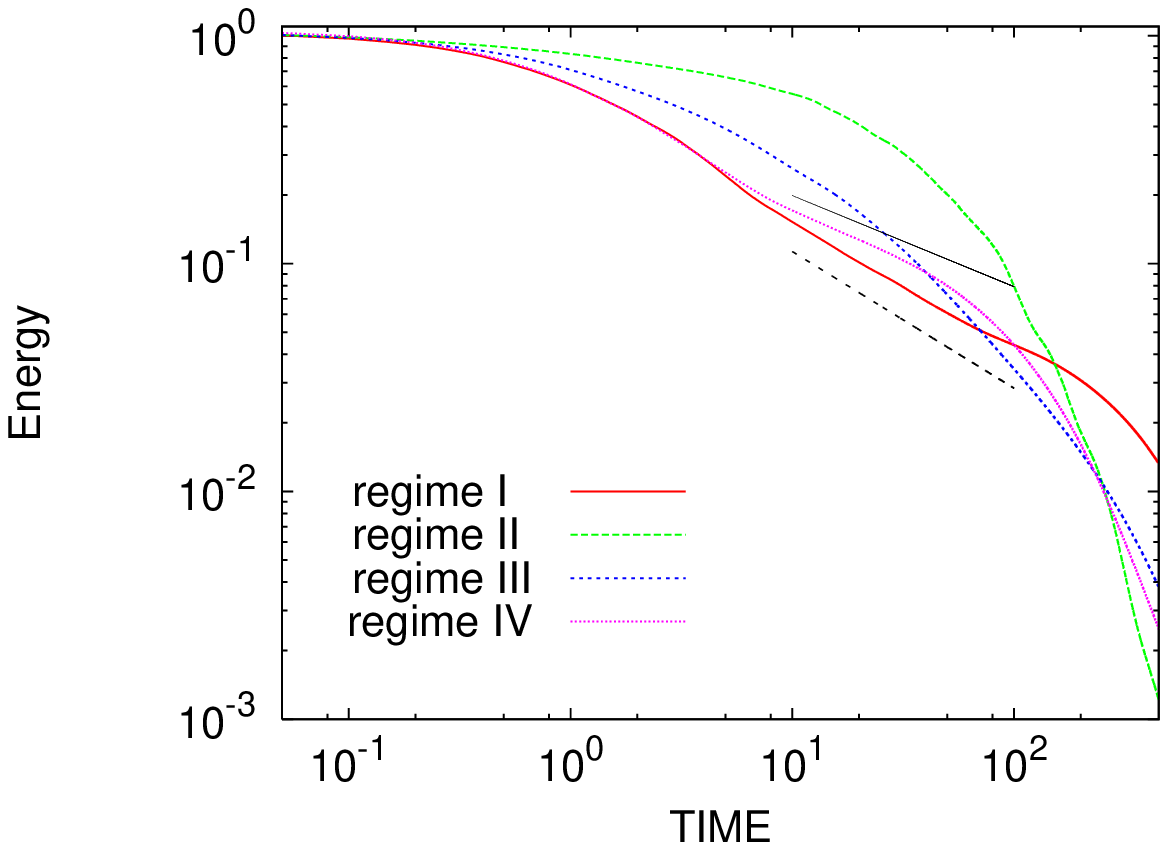}
\includegraphics[width=0.4\unitlength]{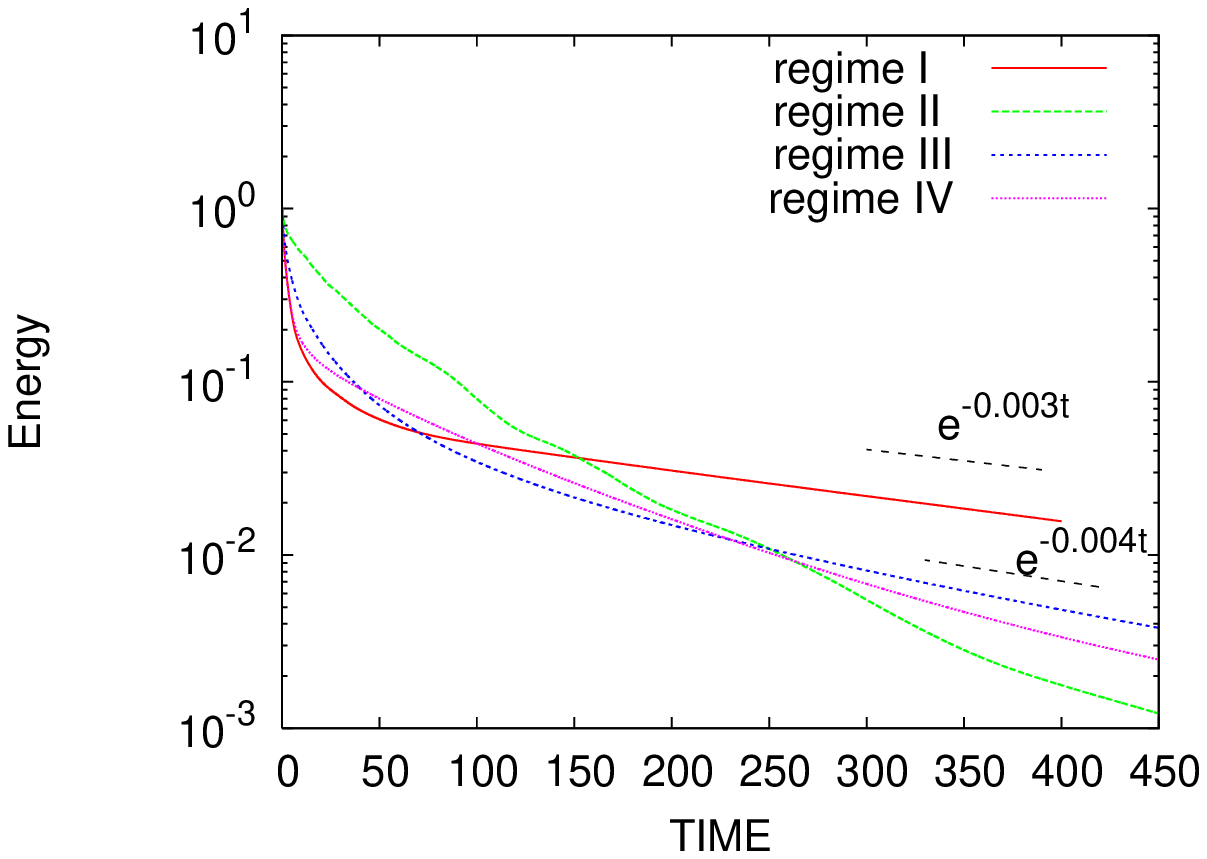}
\caption{Time evolution of the total energy in log-log scale (top) and in log-lin scale (bottom). The solid line (top) corresponds to $t^{-0.4}$ and the dotted line (top) corresponds to $t^{-0.6}$. \label{Eloglin}}
\end{figure}

The decay of total energy is shown in figure \ref{Eloglin}. At intermediate times, the energy in case I and II decays following a powerlaw with exponents varying for the different sets of initial conditions (Fig.~\ref{Eloglin}, top). The exponents of these powerlaws are approximately $-0.6$ (dotted line) for regime I and $-0.4$ (solid line) for regime IV. It is seen that these powerlaws are observed only after an initial period of rapid decay. In the other cases no clear powerlaw behavior can be identified. This can be compared to previous studies \cite{Kinney1995,Biskamp2001} in which values around $-0.75$ and $-1$ were found for the decaying periodic case. In case III, in which dynamic alignment is observed, no clear power-law behavior is observed. In this case the nonlinear interactions are progressively damped by the alignment process, so that no selfsimilar period is observed in the energy decay. At late times (Fig. \ref{Eloglin}, bottom) all cases show an exponential viscous decay of the form $E\sim e^{-2\alpha\nu t}$ with $\alpha=1.5$ in case I and $\alpha=2$ in cases II, III and IV, a value related to the largest Stokes eigenmode of the circle ($\alpha=1.64$), which contains most of the energy, as found in \cite{Schneider2007} for the hydrodynamical case.

\begin{figure*}
\begin{center}
 \includegraphics[scale=0.2]{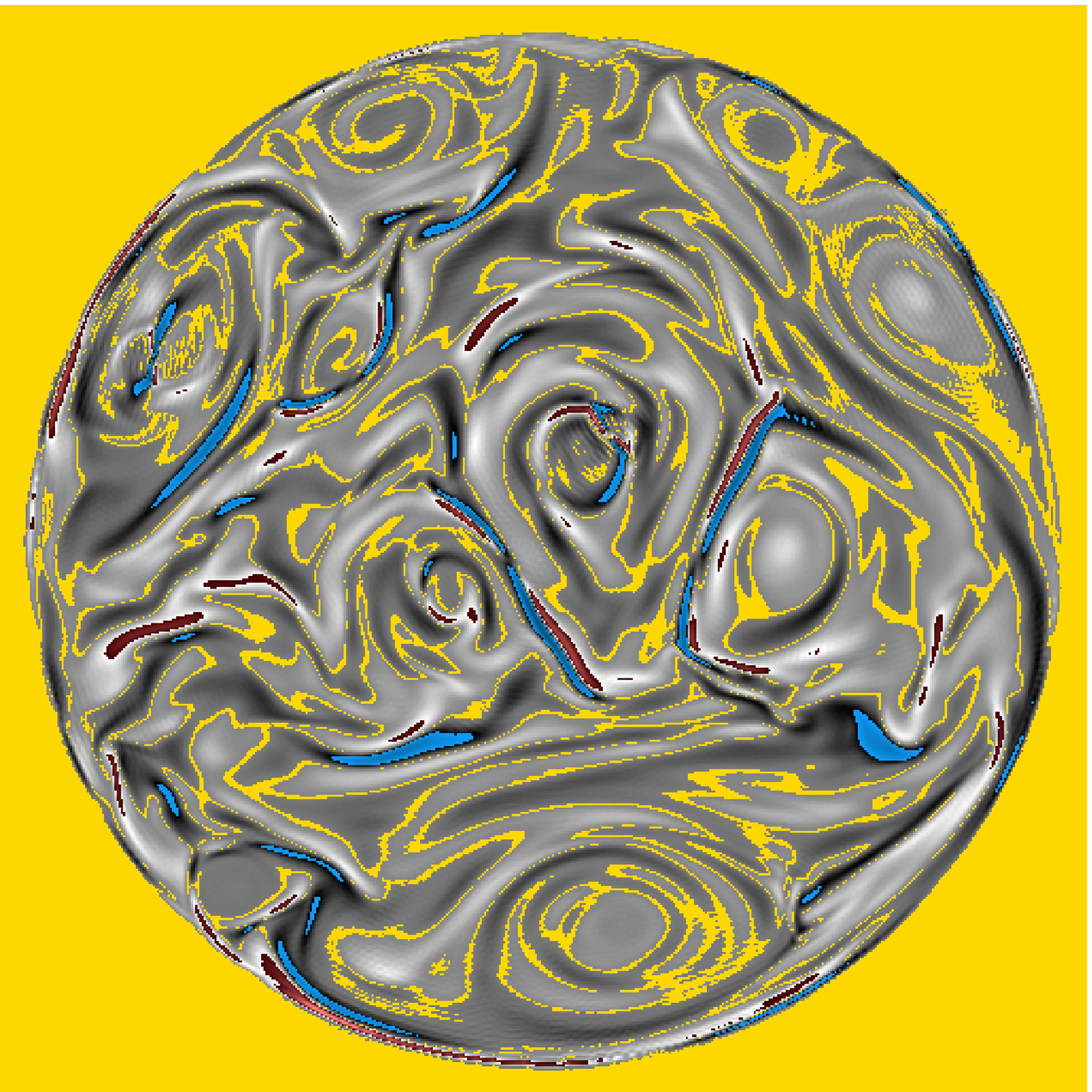}
 \includegraphics[scale=0.2]{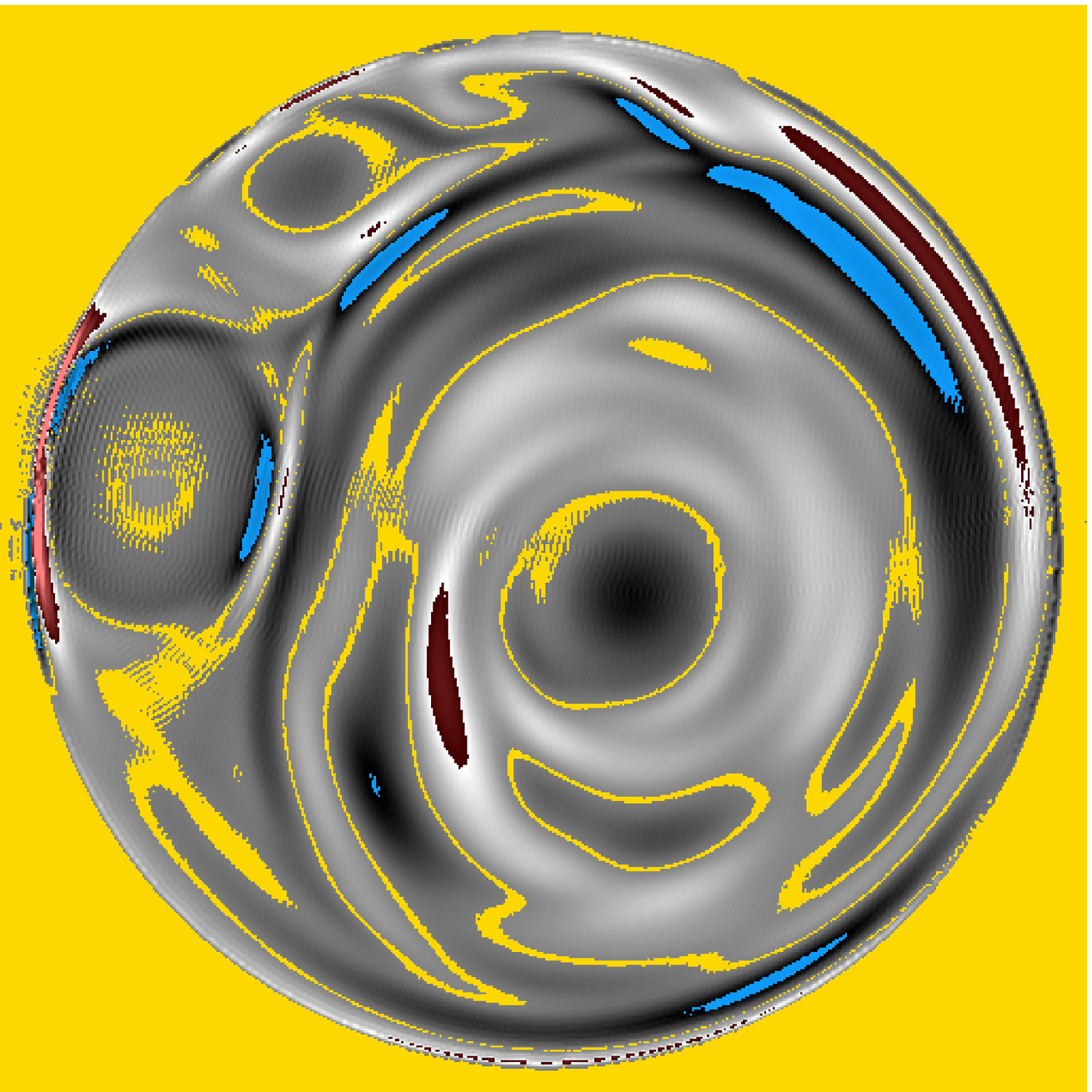}
 \includegraphics[scale=0.2]{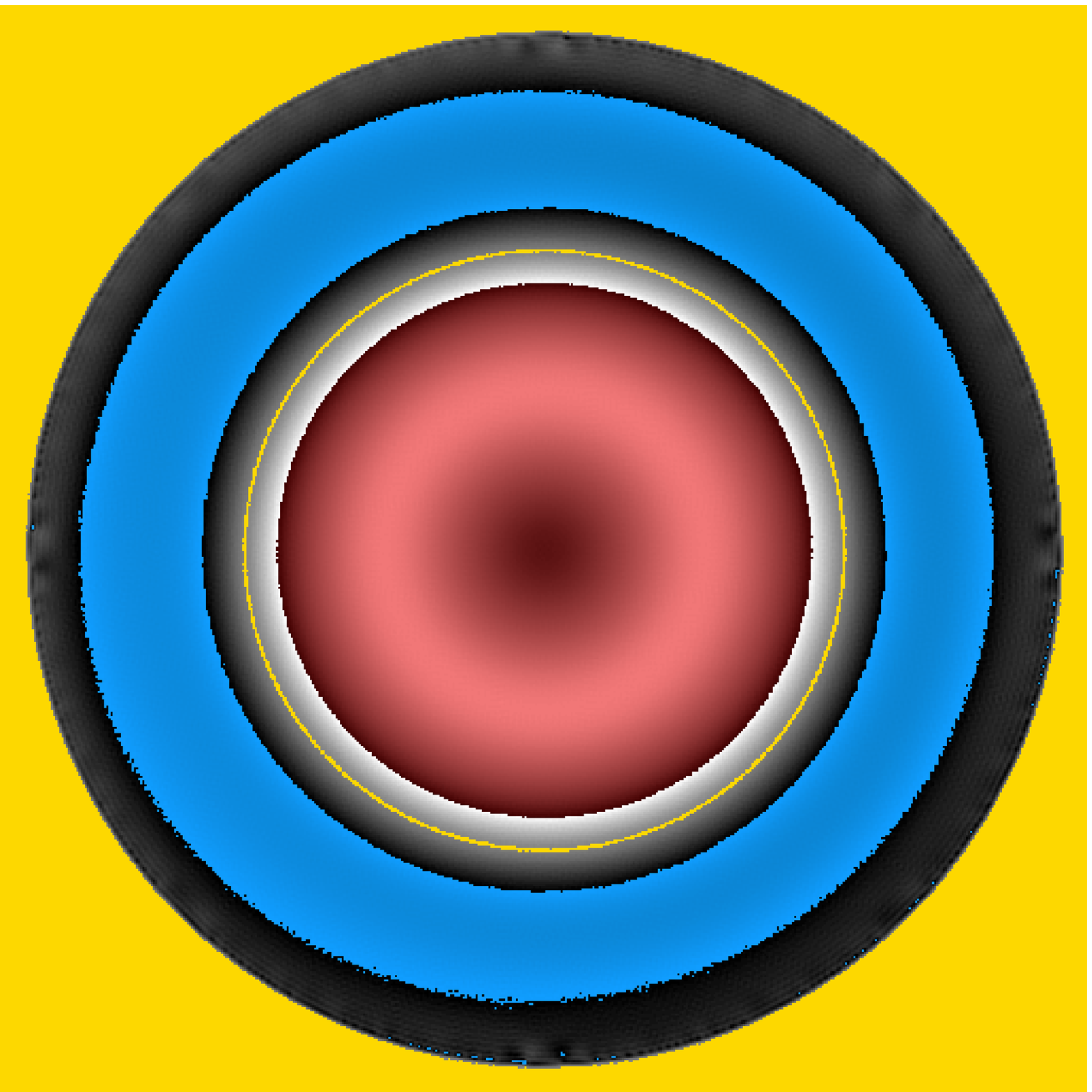}\\
 \includegraphics[scale=0.2]{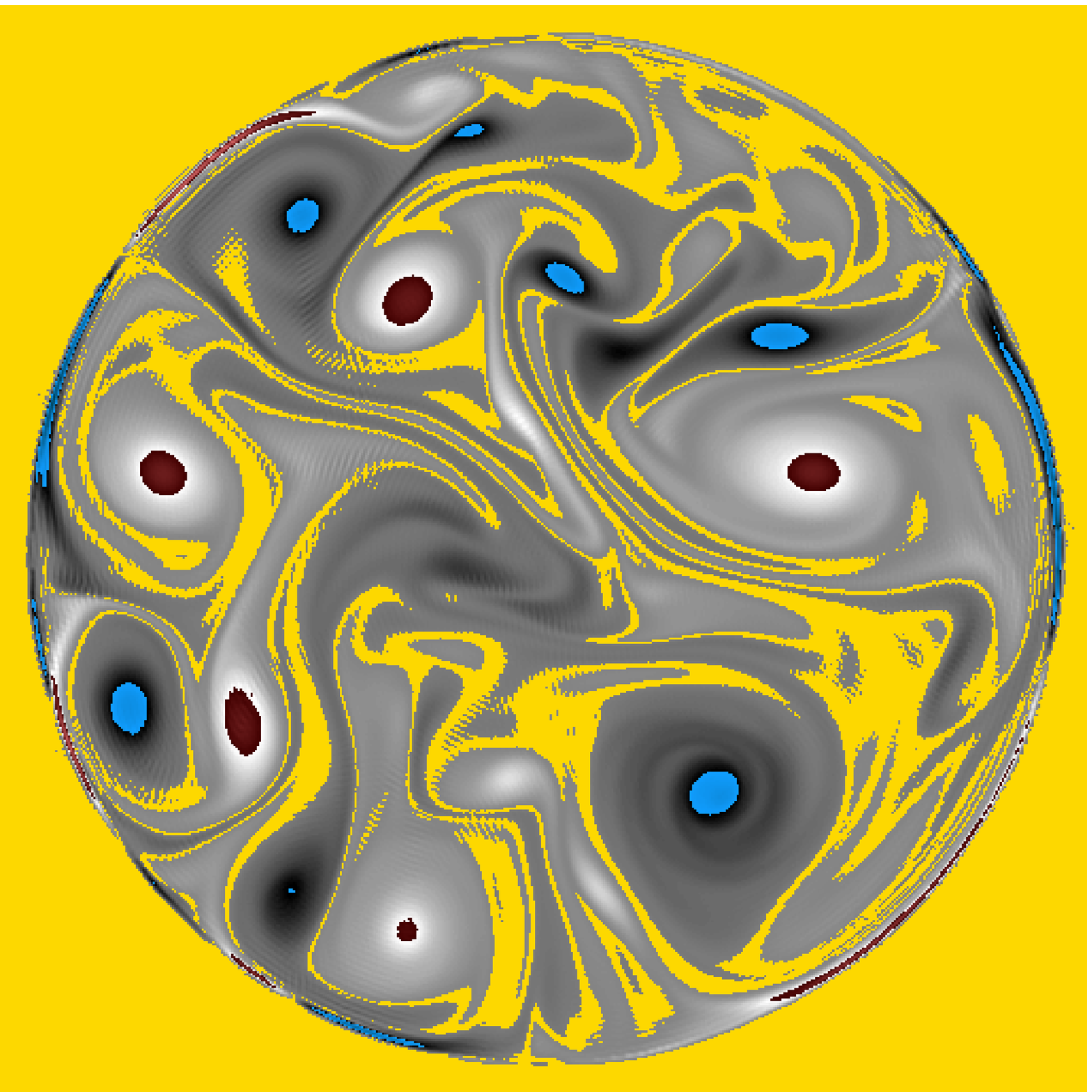}
 \includegraphics[scale=0.2]{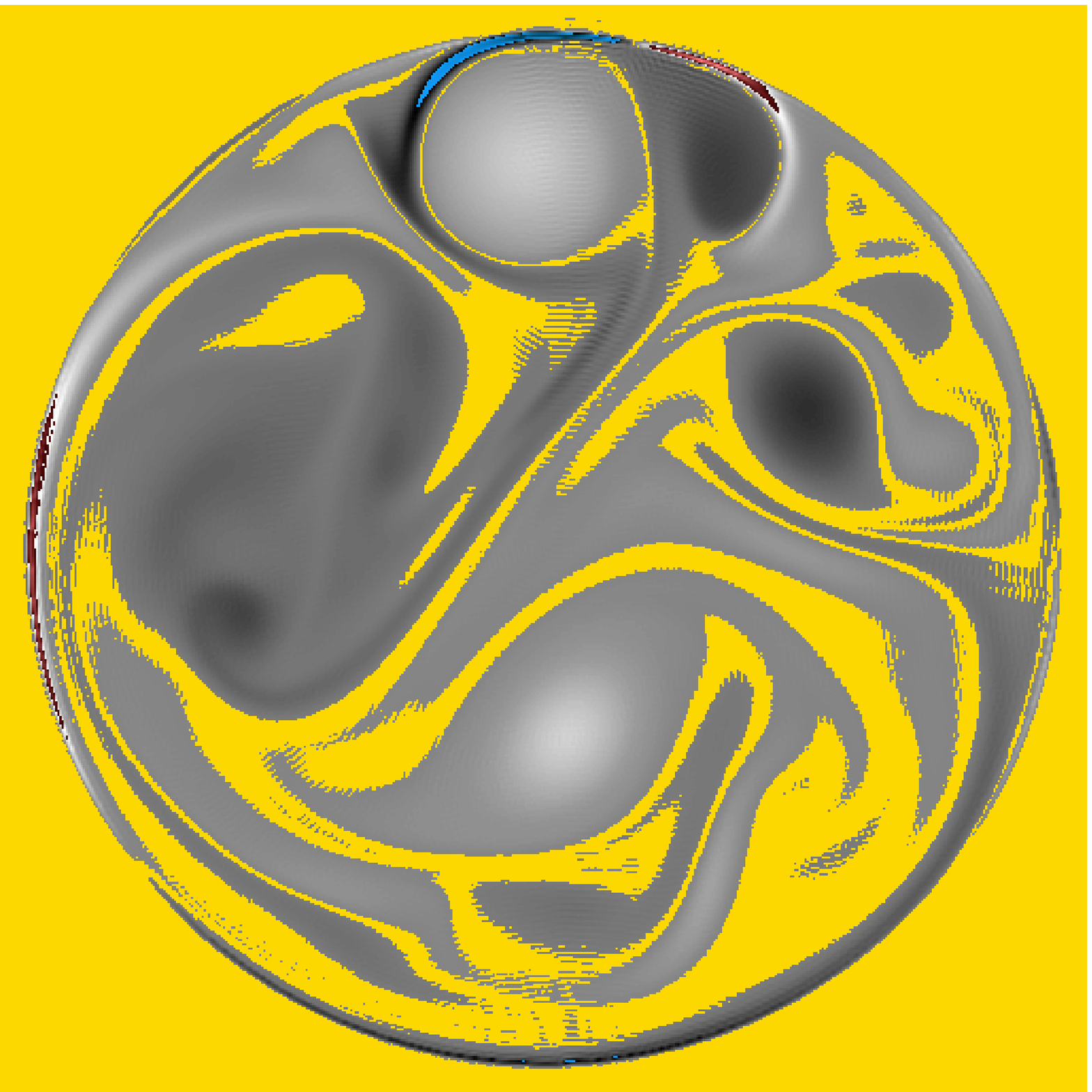}
 \includegraphics[scale=0.2]{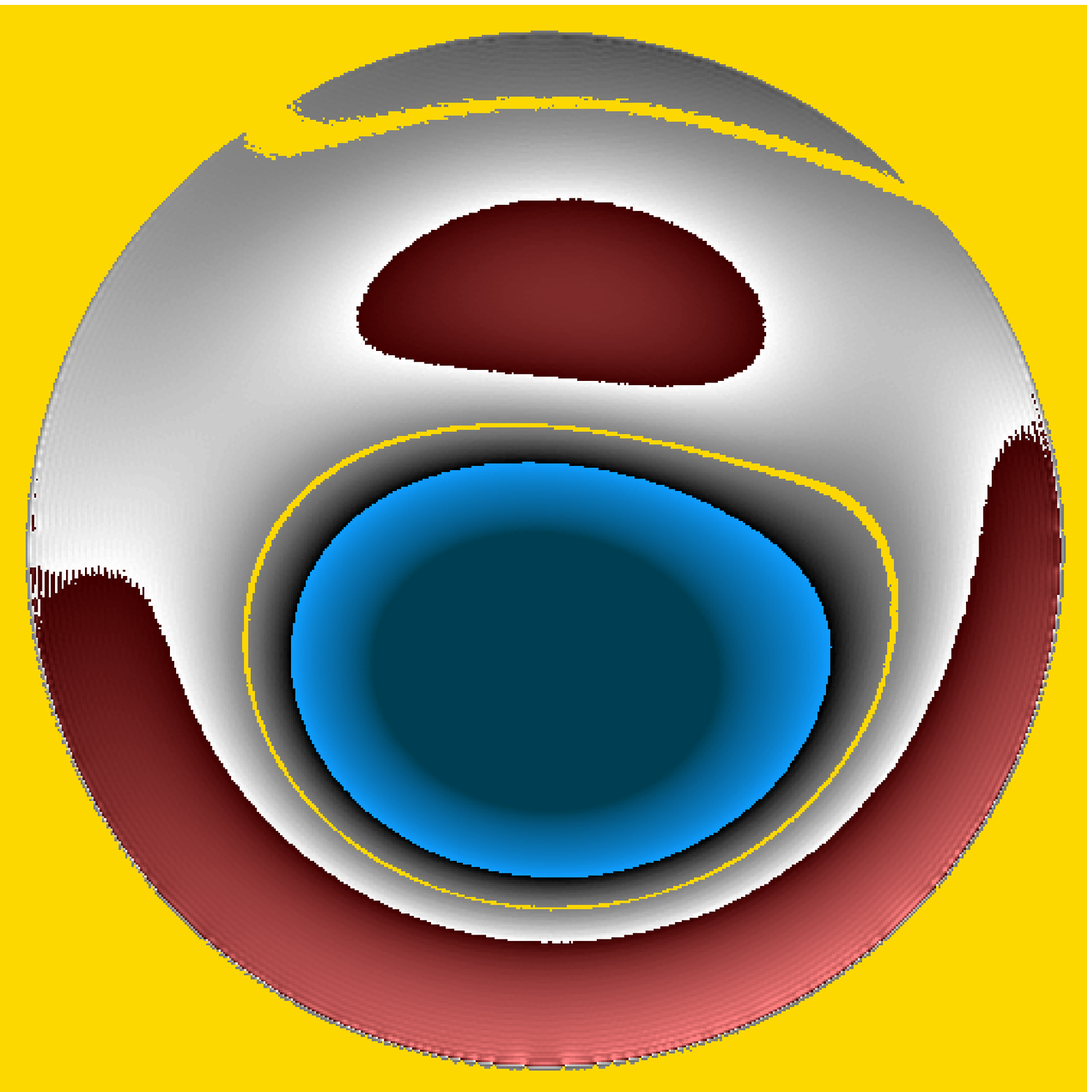}\\
 \includegraphics[scale=0.2]{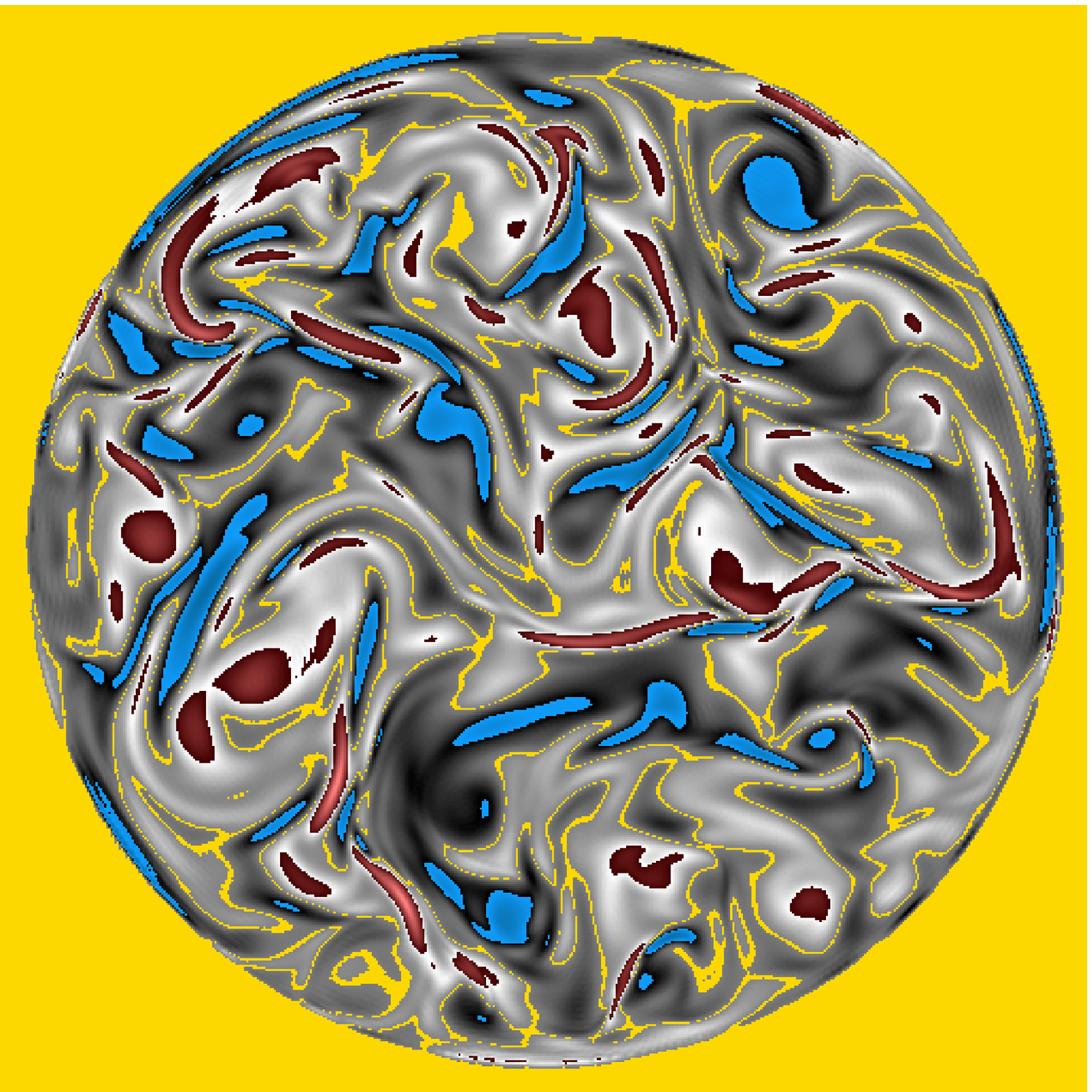}
 \includegraphics[scale=0.2]{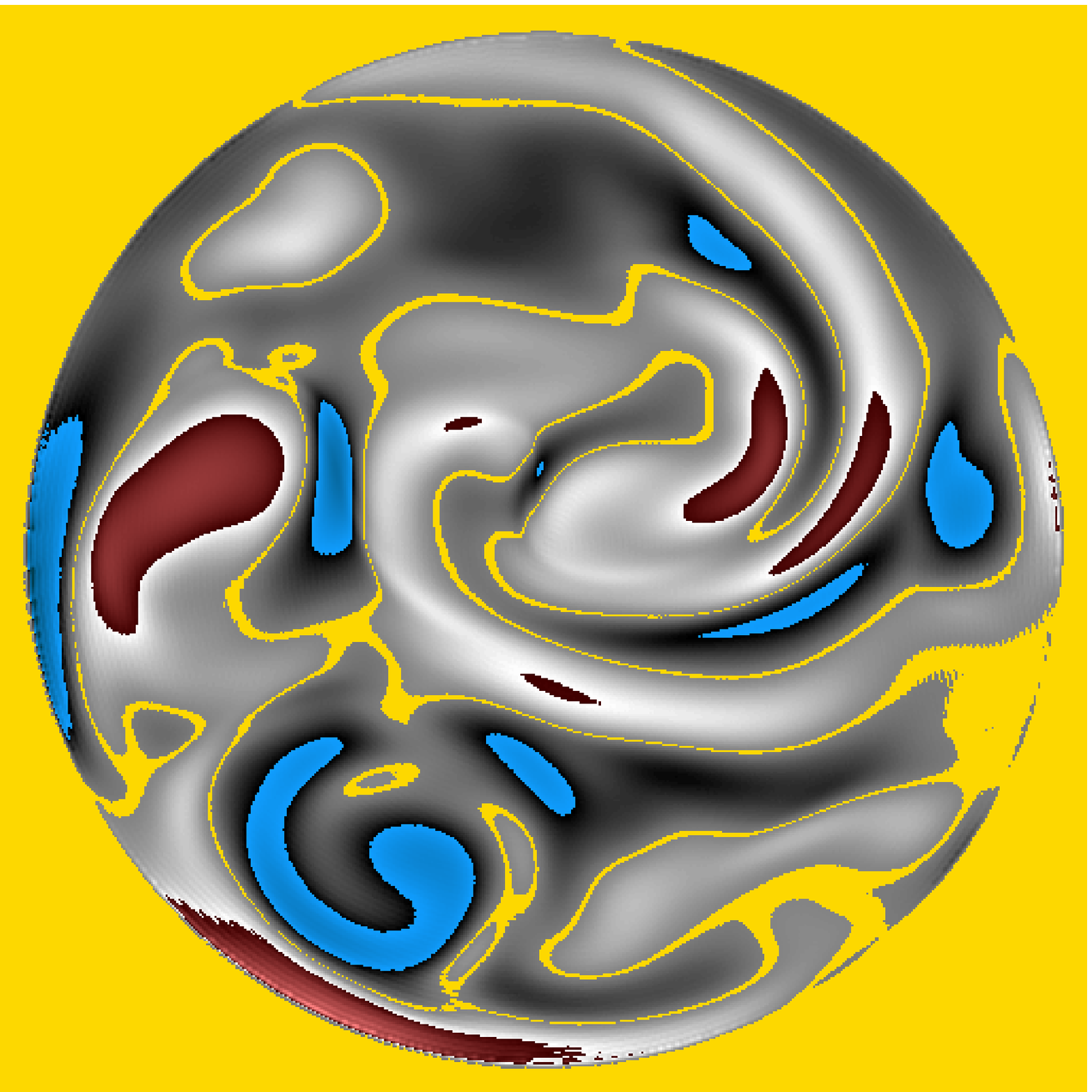}
 \includegraphics[scale=0.2]{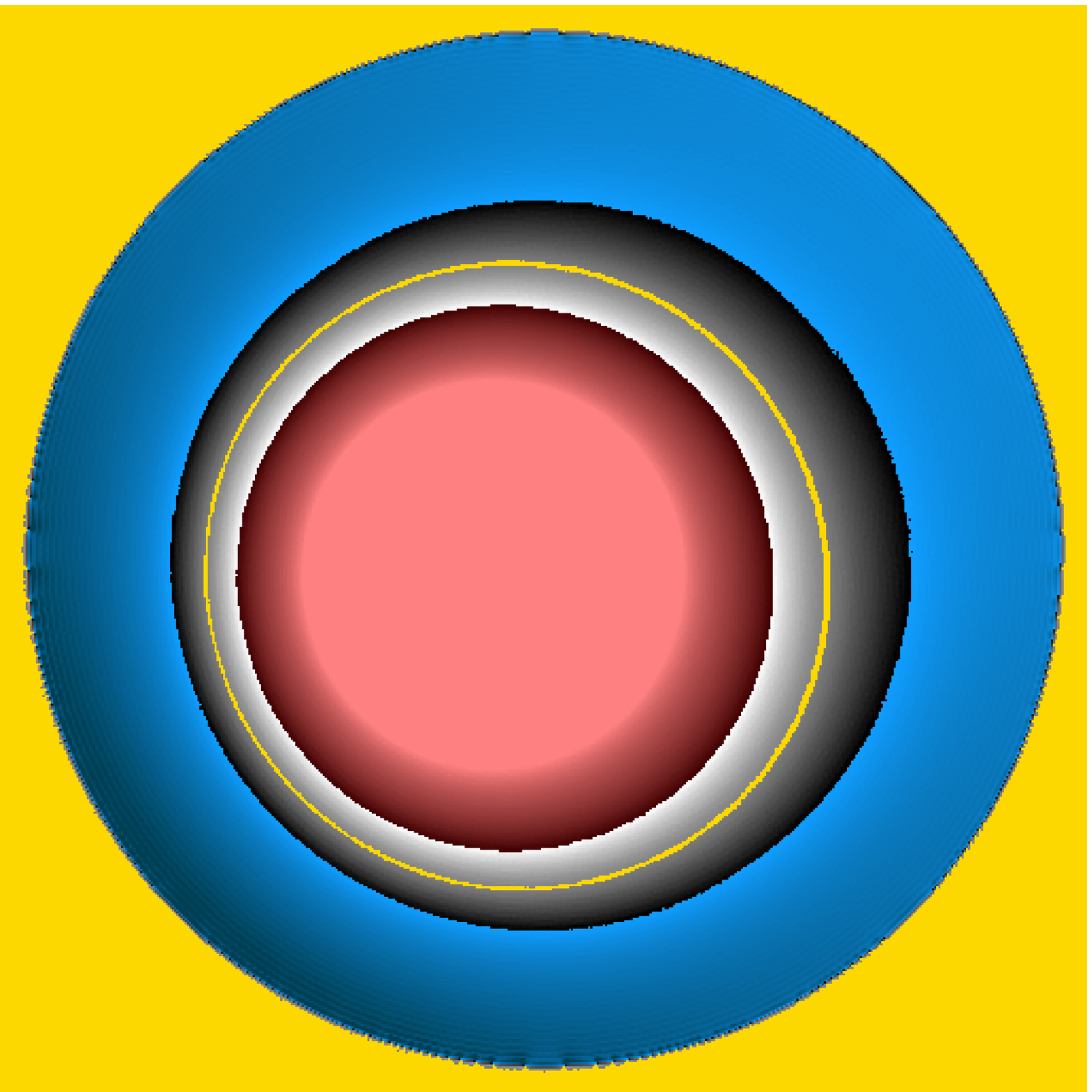}\\
 \includegraphics[scale=0.2]{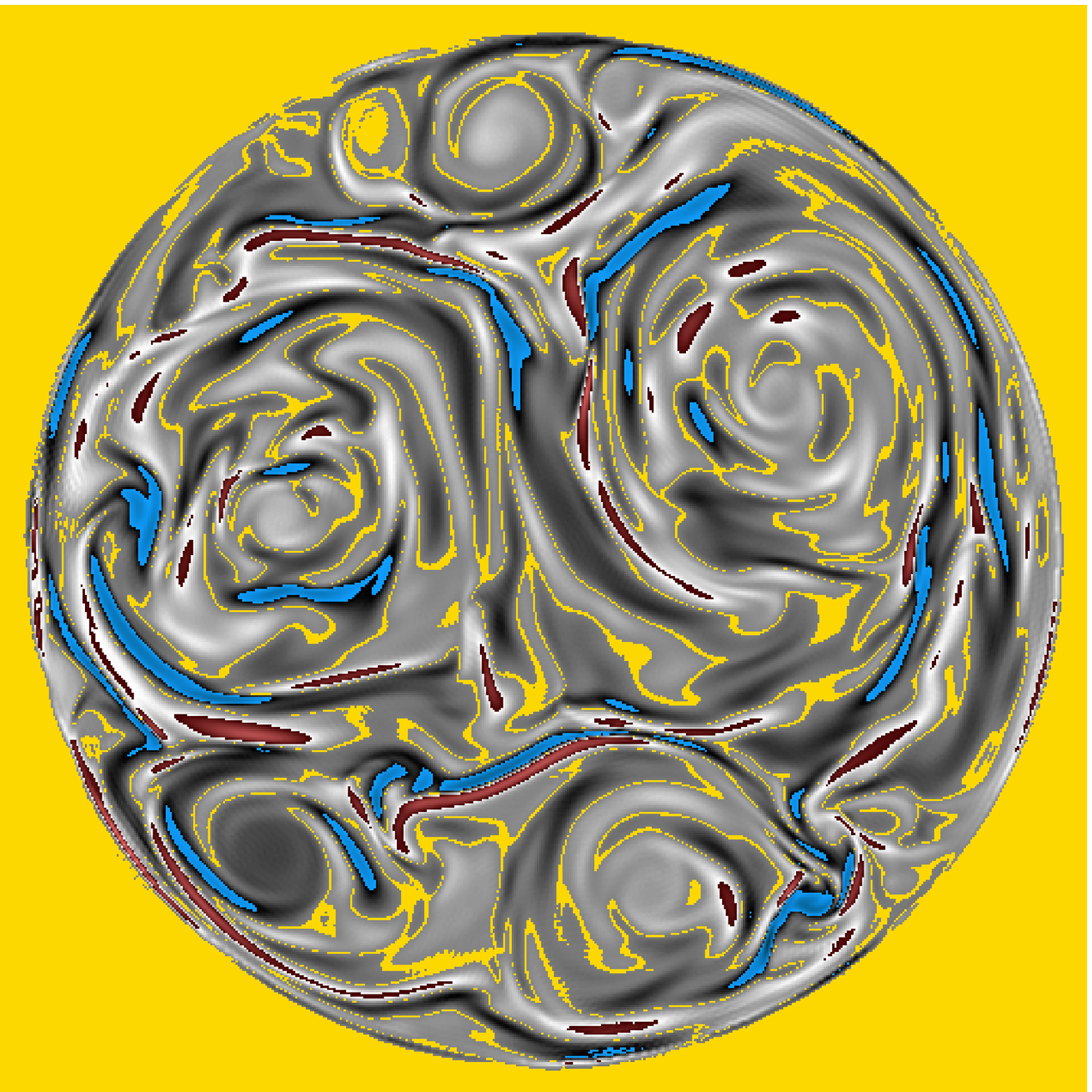}
 \includegraphics[scale=0.2]{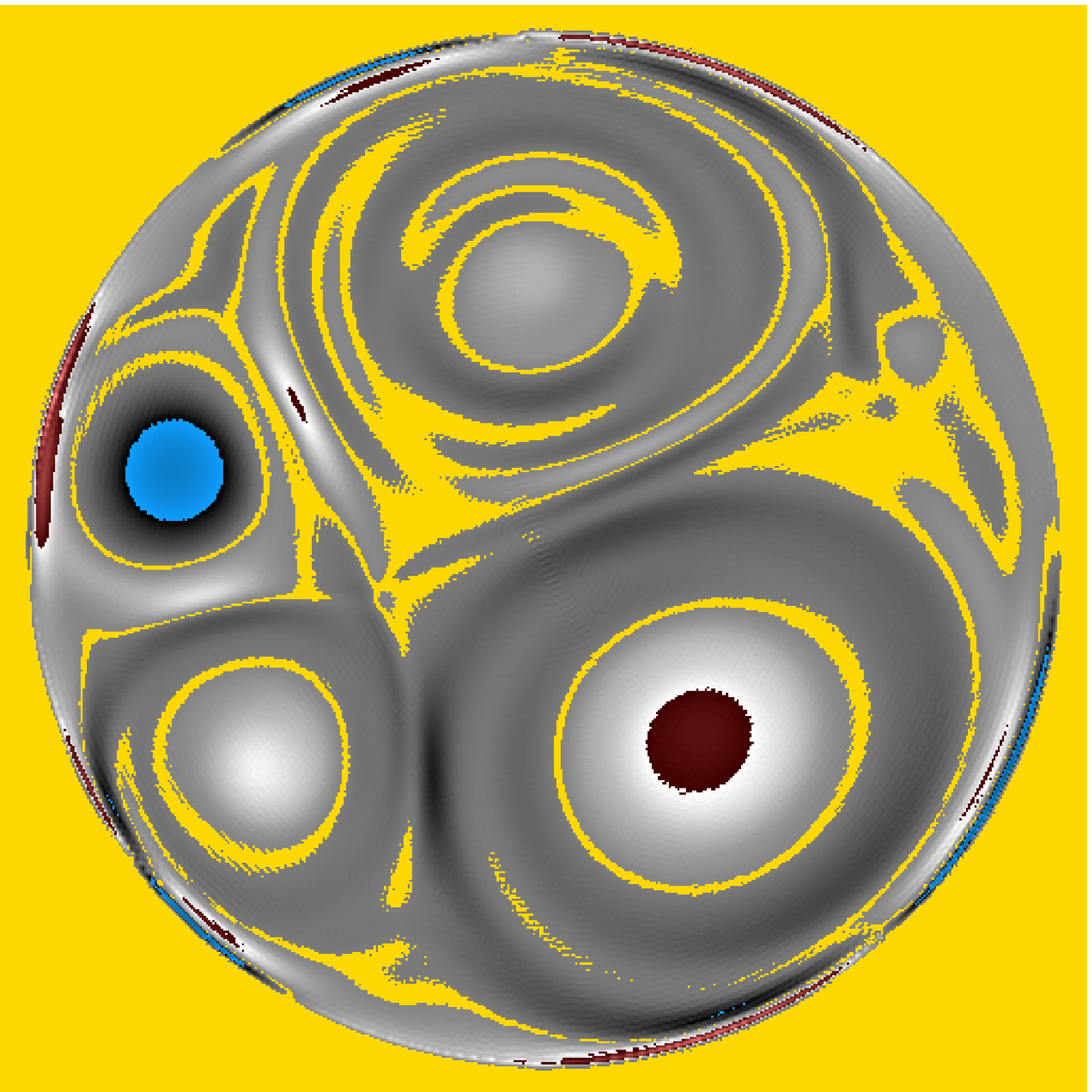}
 \includegraphics[scale=0.2]{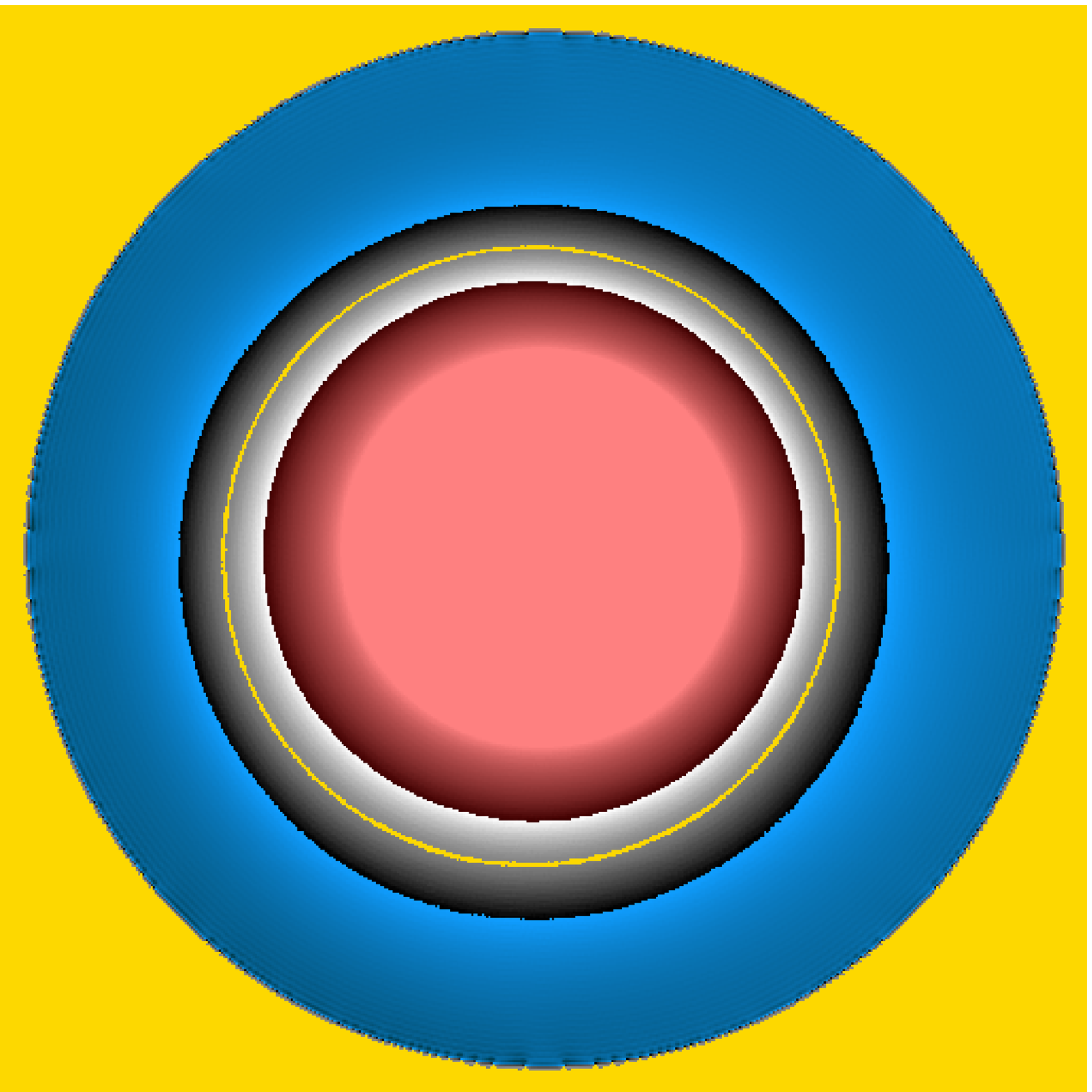}

 \includegraphics[scale=0.4,angle=270]{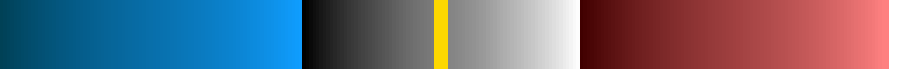}
\end{center}
\caption{\label{vort.per}\footnotesize{Vorticity at different instants in the circular domain. From top to bottom: regime I, regime II, regime III and regime IV; from left to right: $t=5$, $t=40$ and in the last column the time coresponds to $t=250$ for regime I and $t=1250$ for regimes II, III and IV.}\label{visuvort}}
\end{figure*}

\begin{figure*}
\begin{center}
 \includegraphics[scale=0.2]{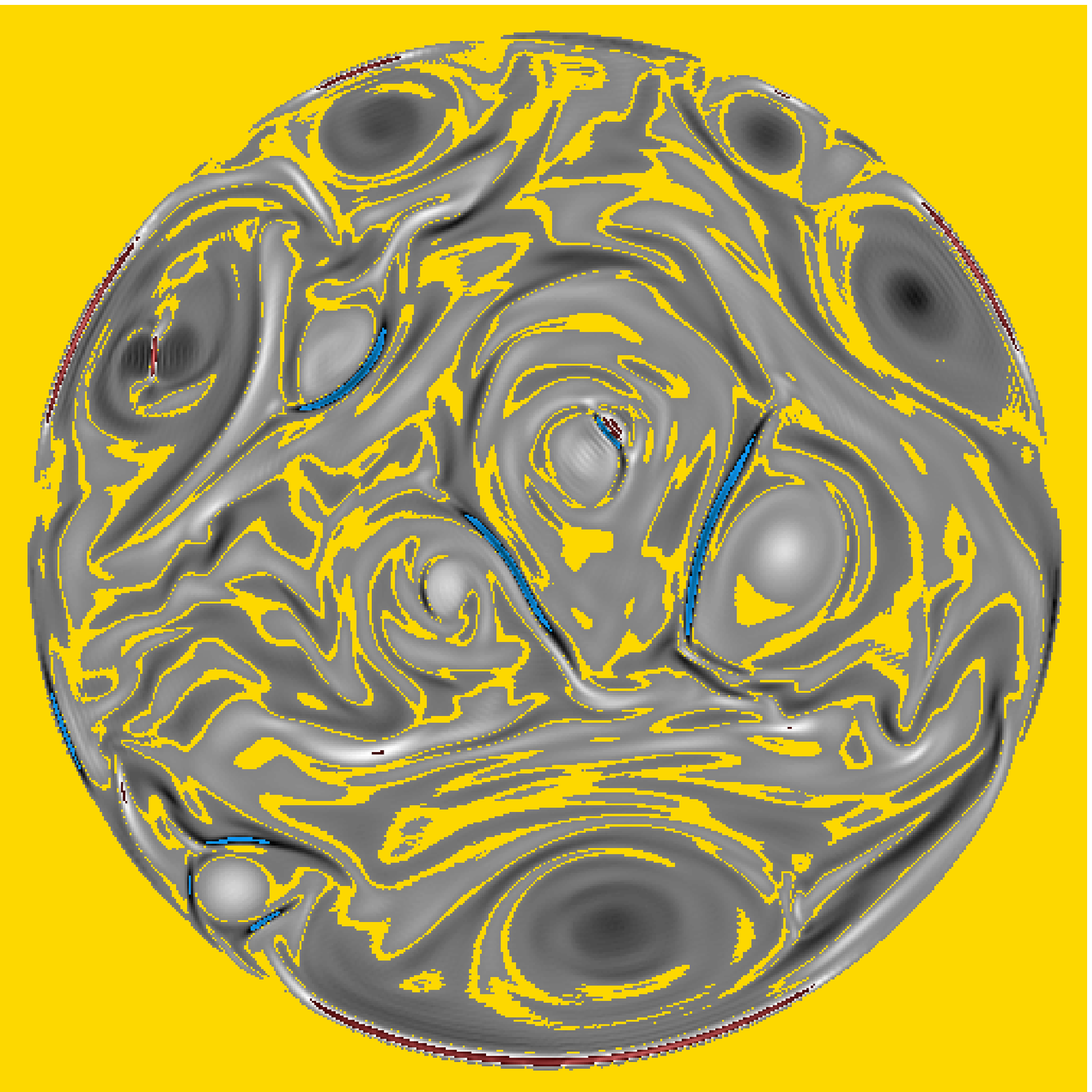}
 \includegraphics[scale=0.2]{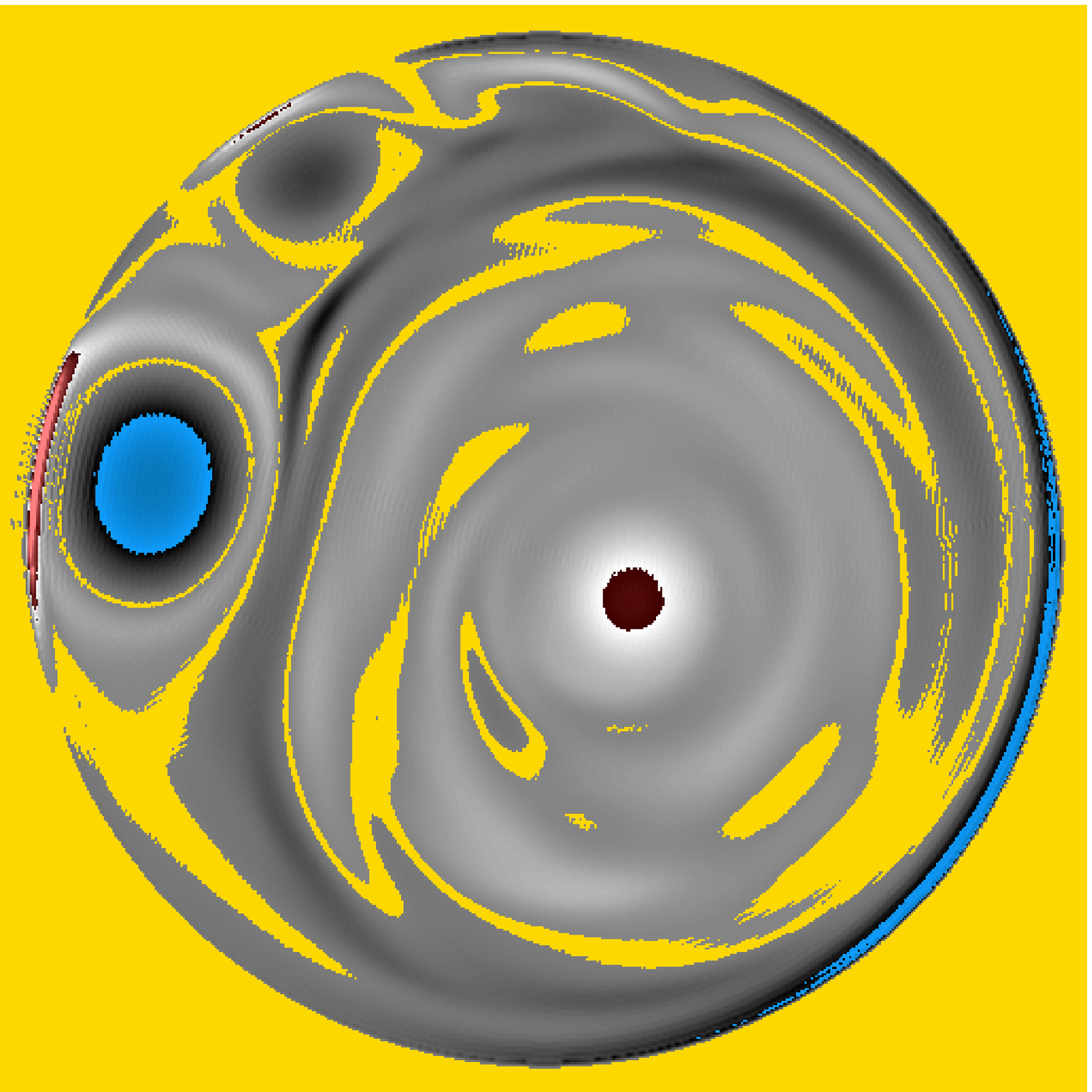}
 \includegraphics[scale=0.2]{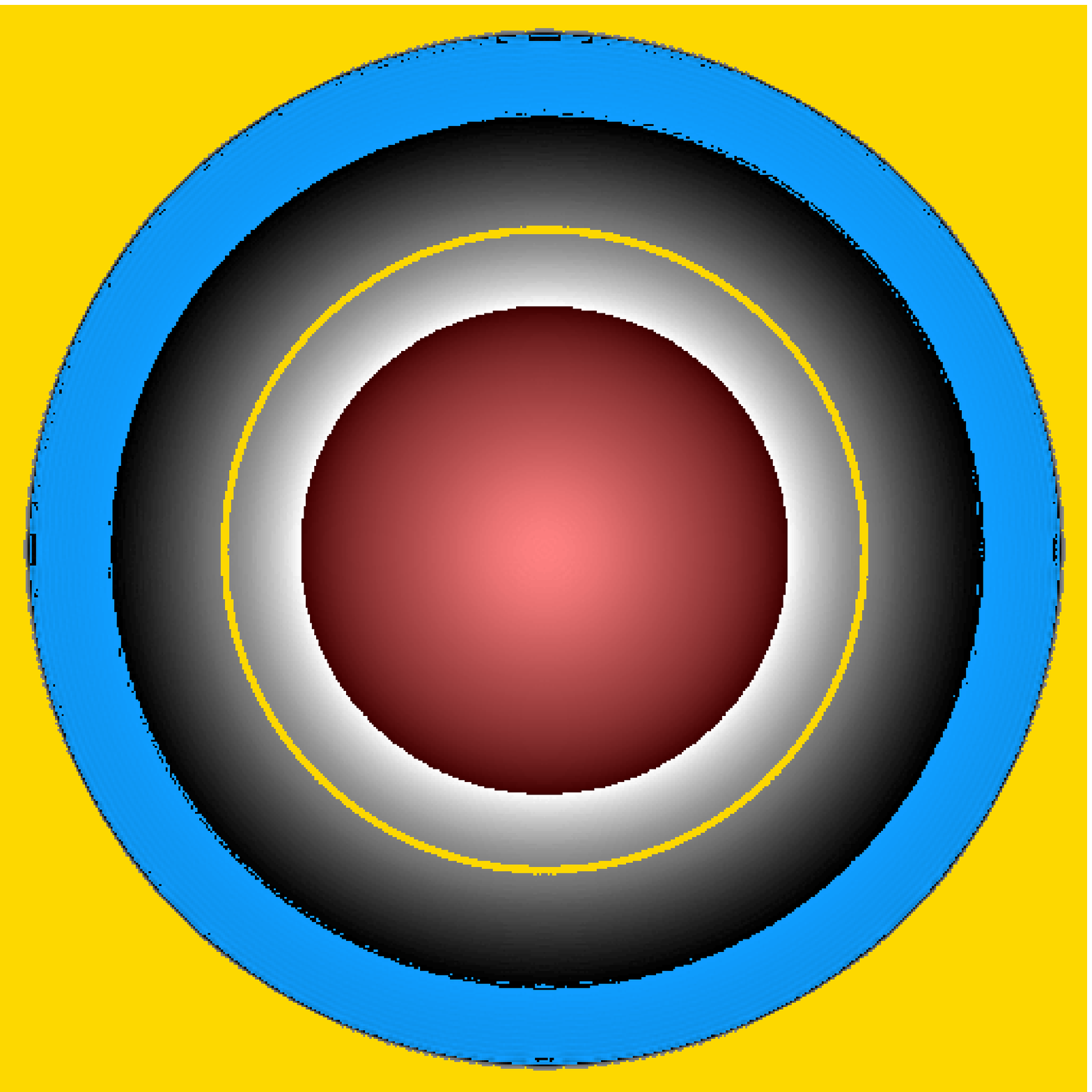}\\
 \includegraphics[scale=0.2]{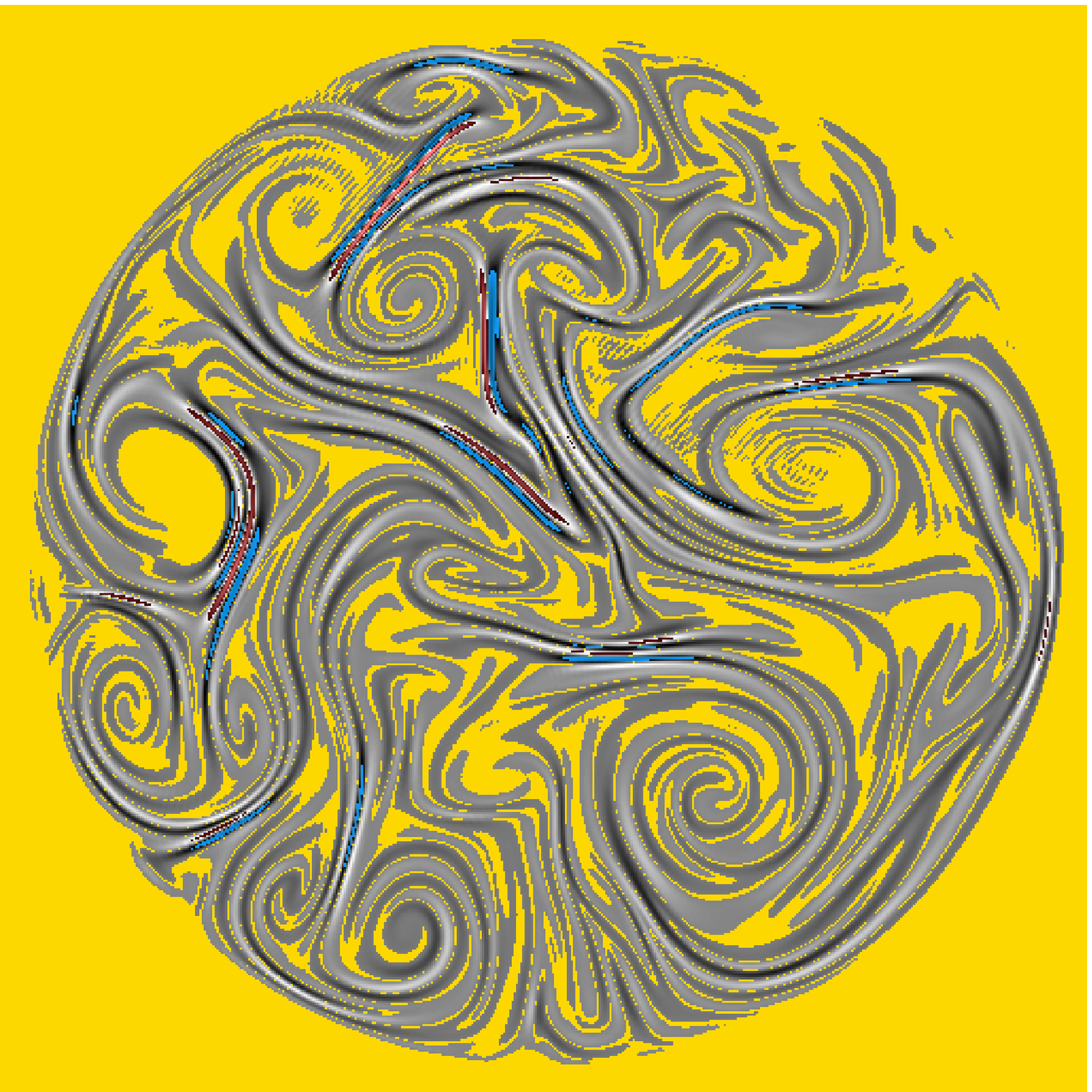}
 \includegraphics[scale=0.2]{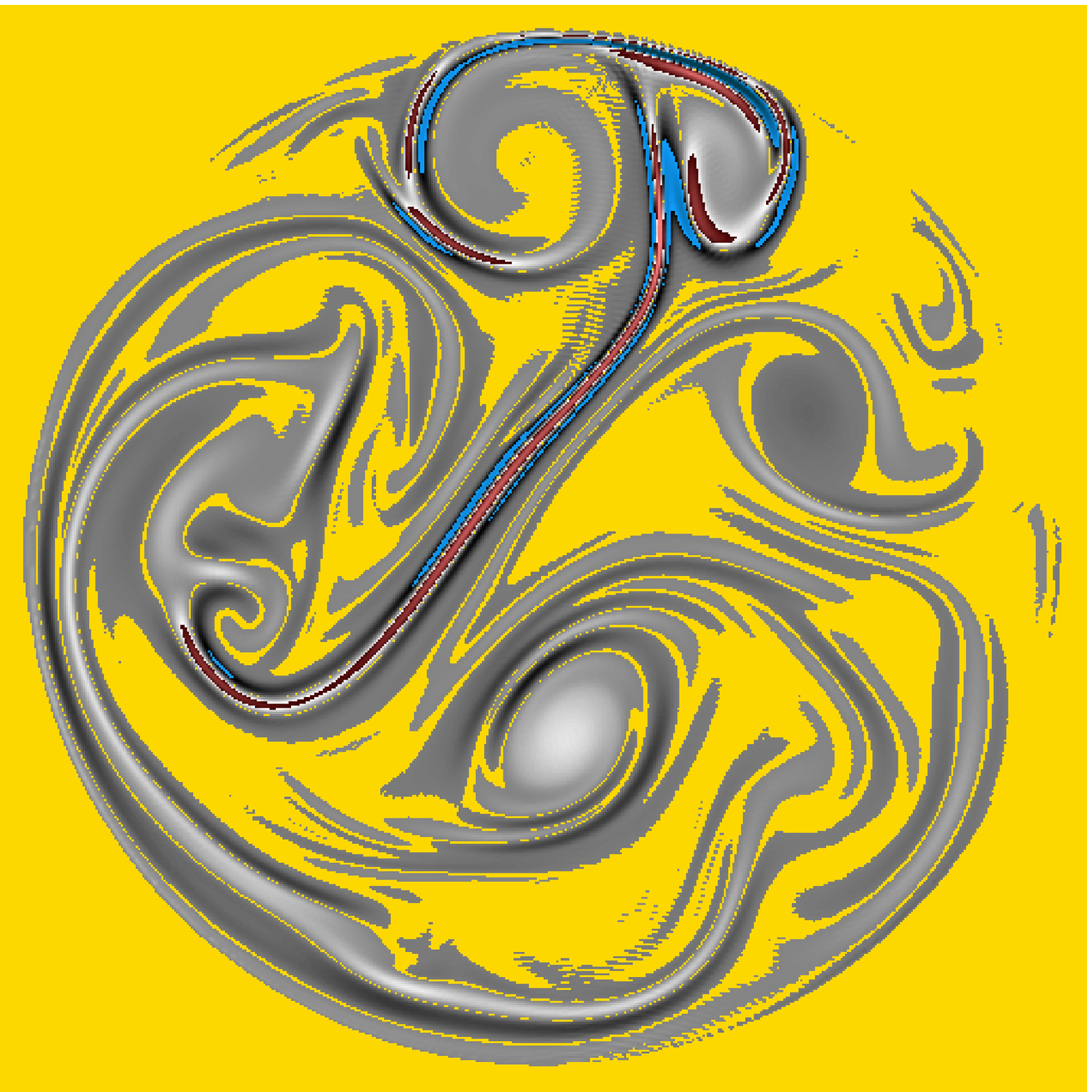}
 \includegraphics[scale=0.2]{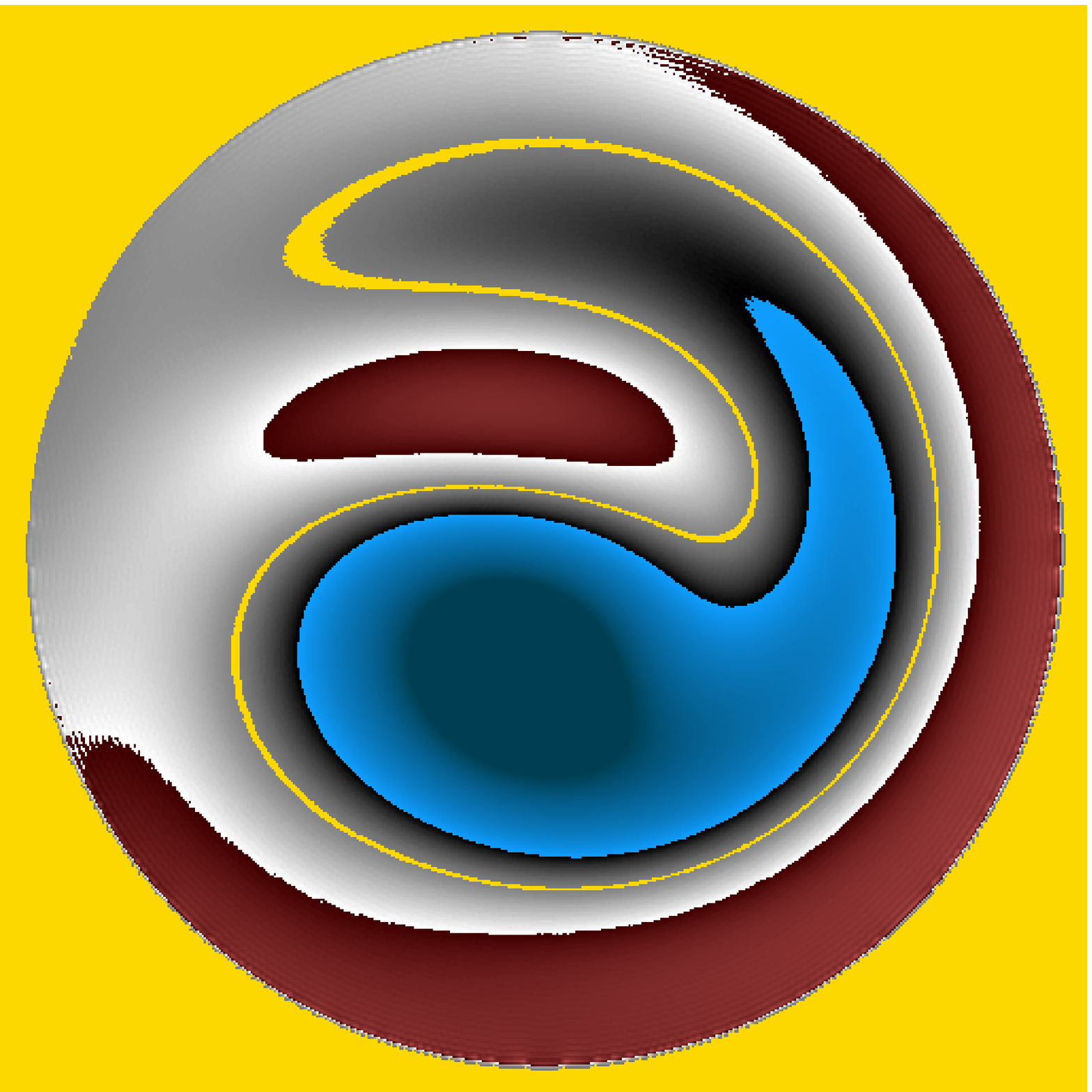}\\
 \includegraphics[scale=0.2]{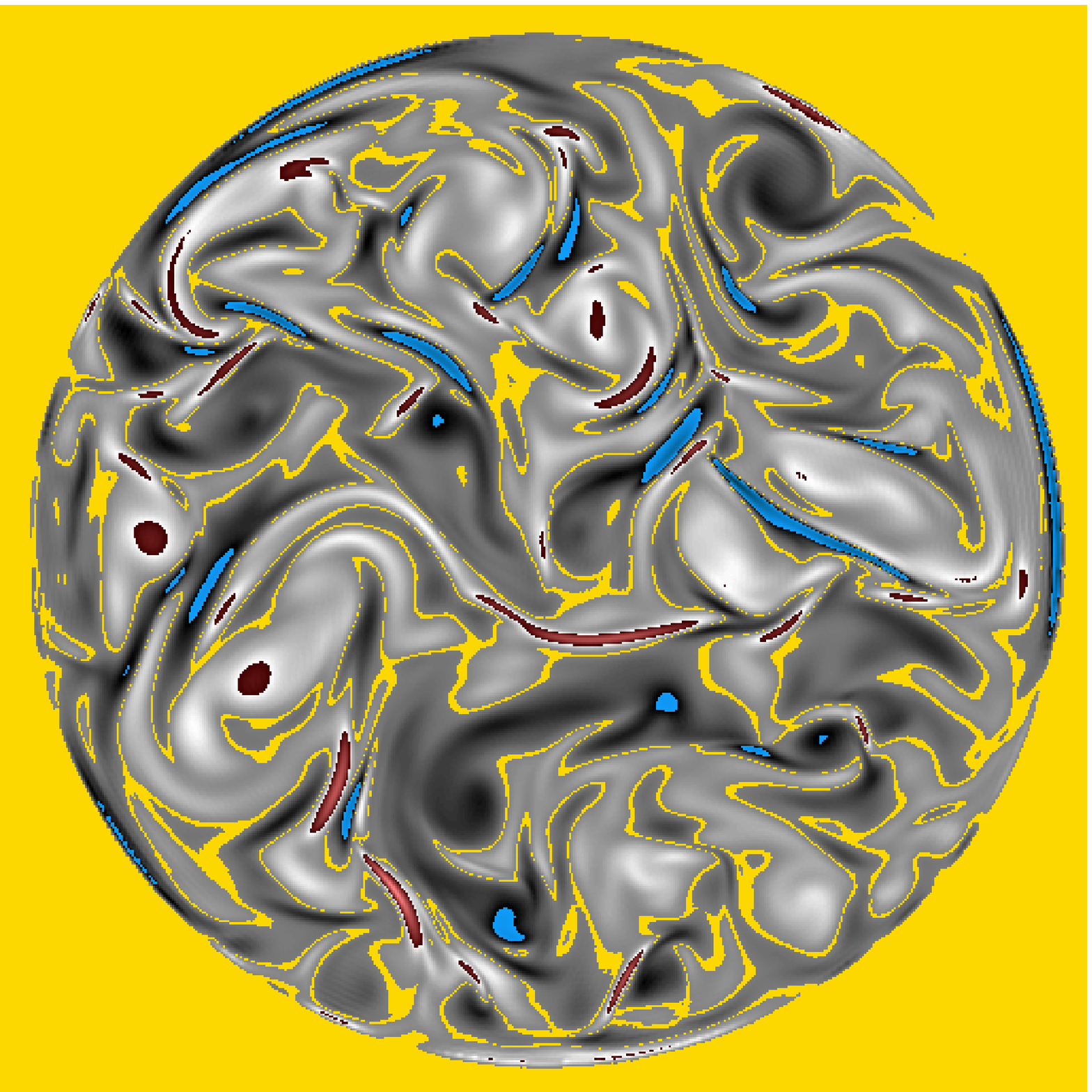}
 \includegraphics[scale=0.2]{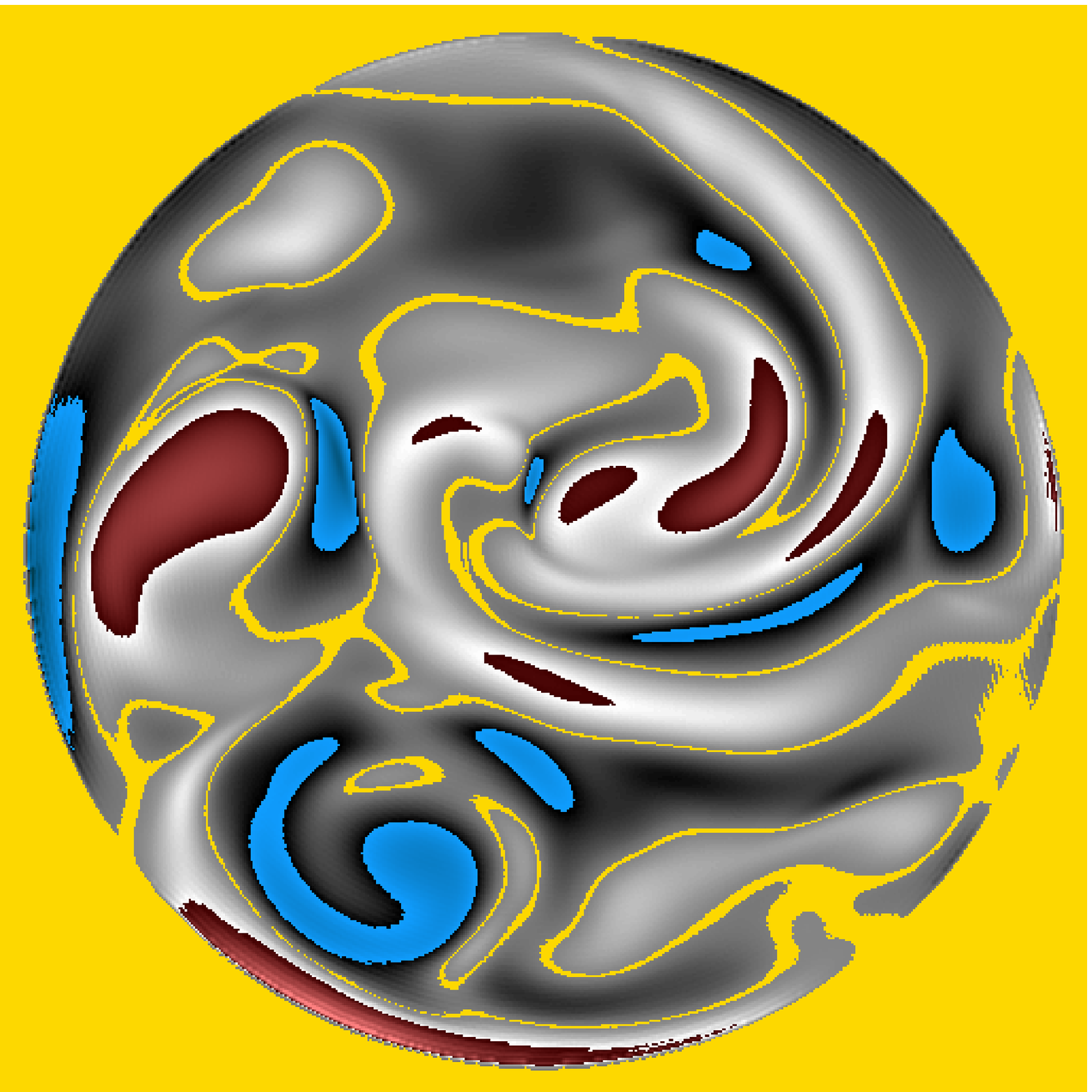}
 \includegraphics[scale=0.2]{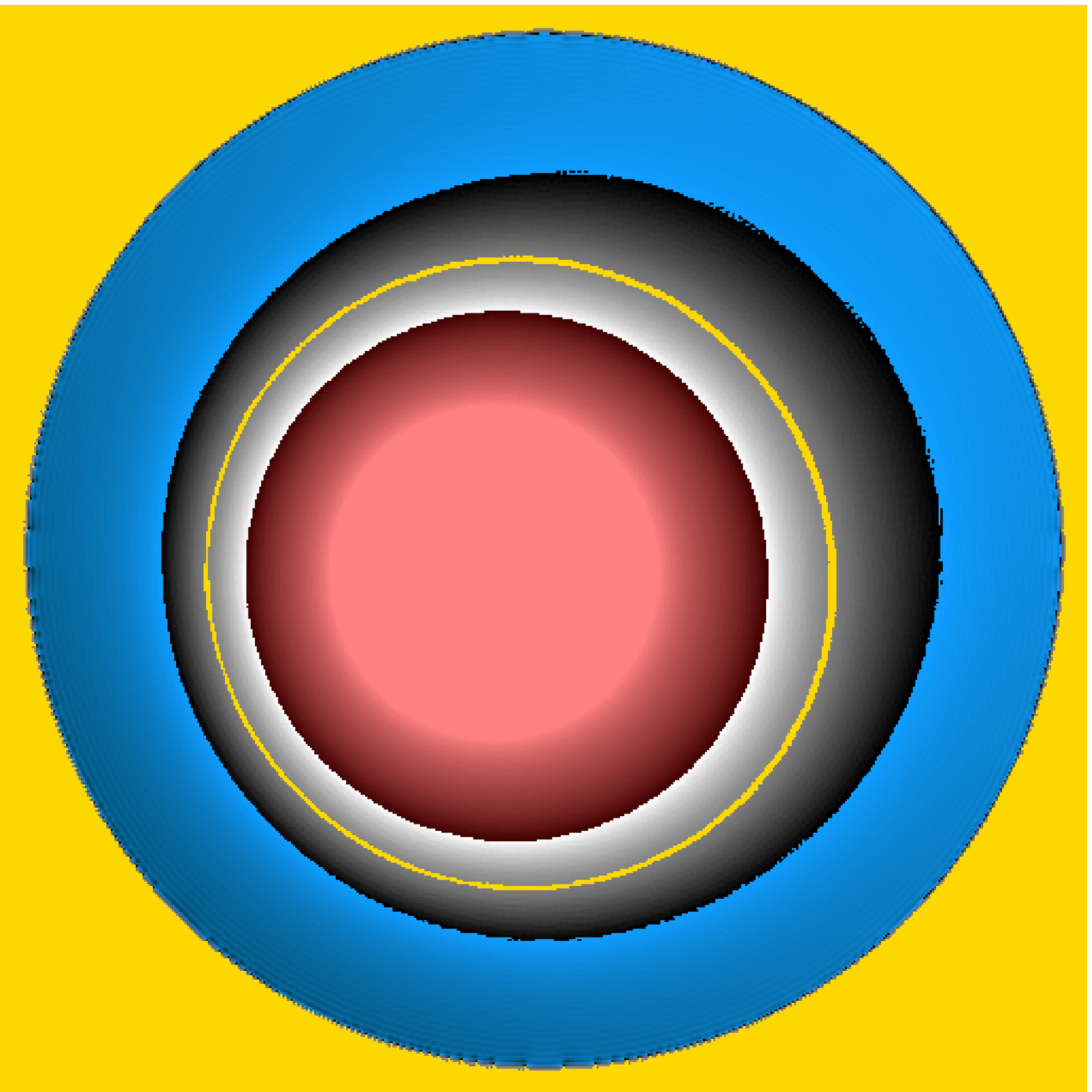}\\
 \includegraphics[scale=0.2]{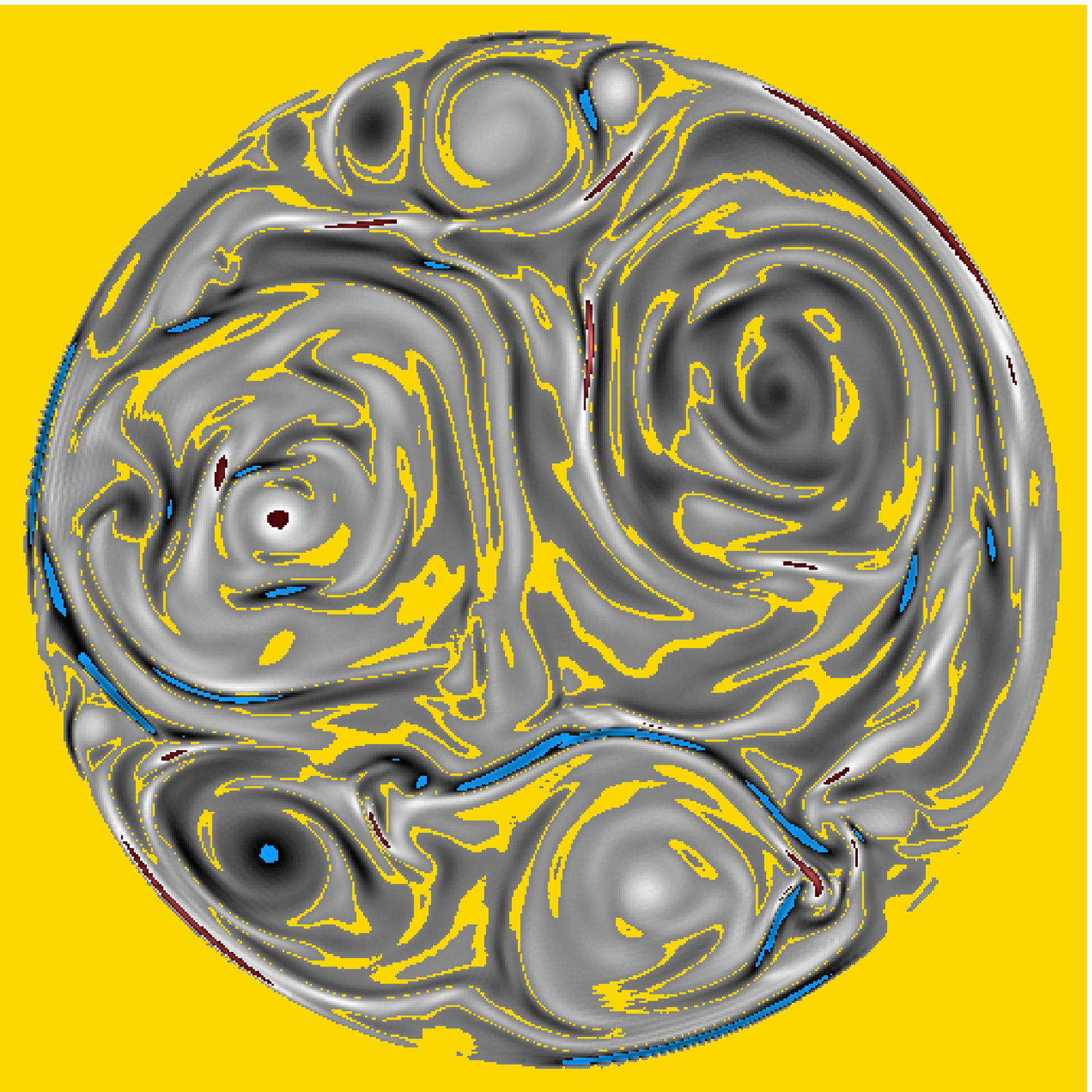}
 \includegraphics[scale=0.2]{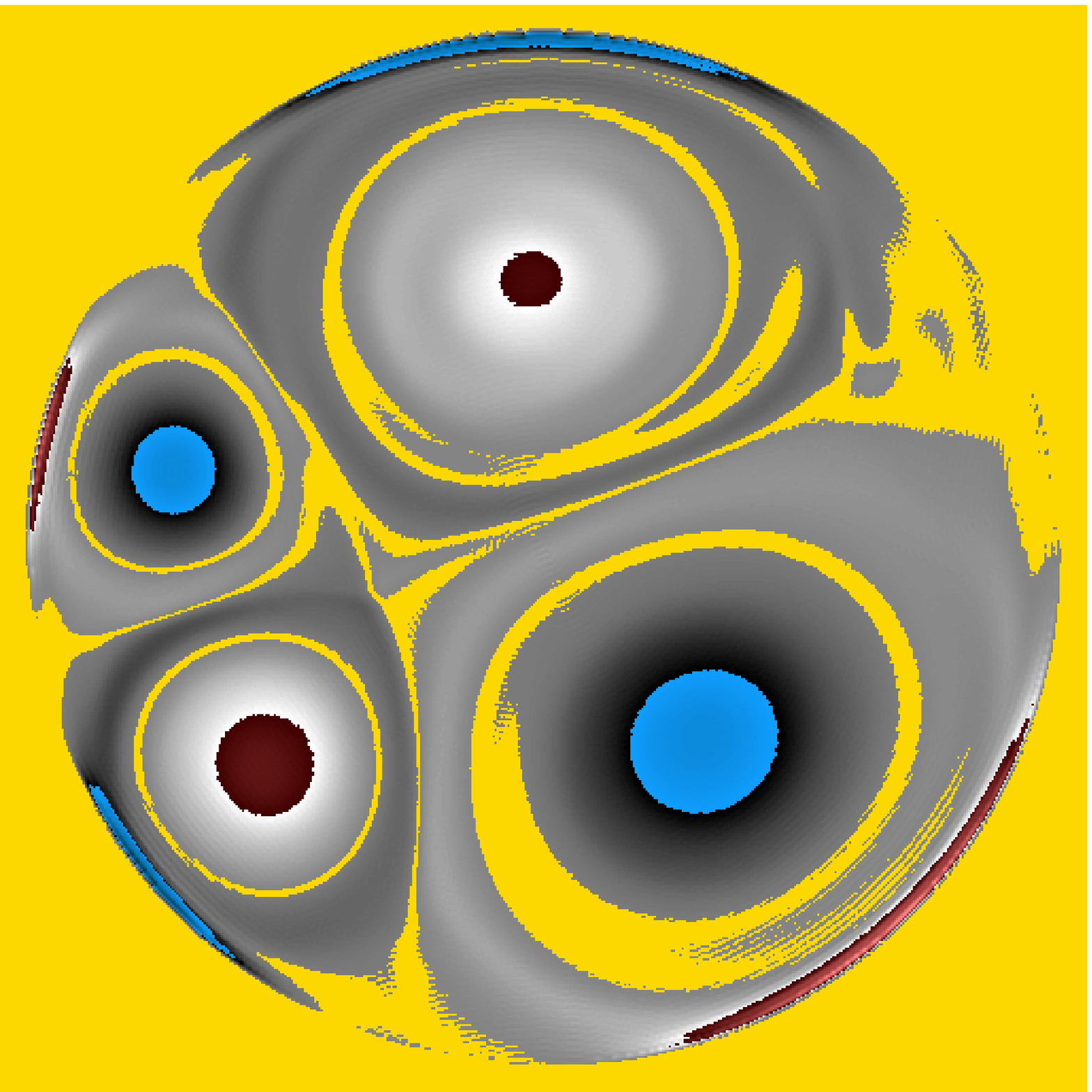}
 \includegraphics[scale=0.2]{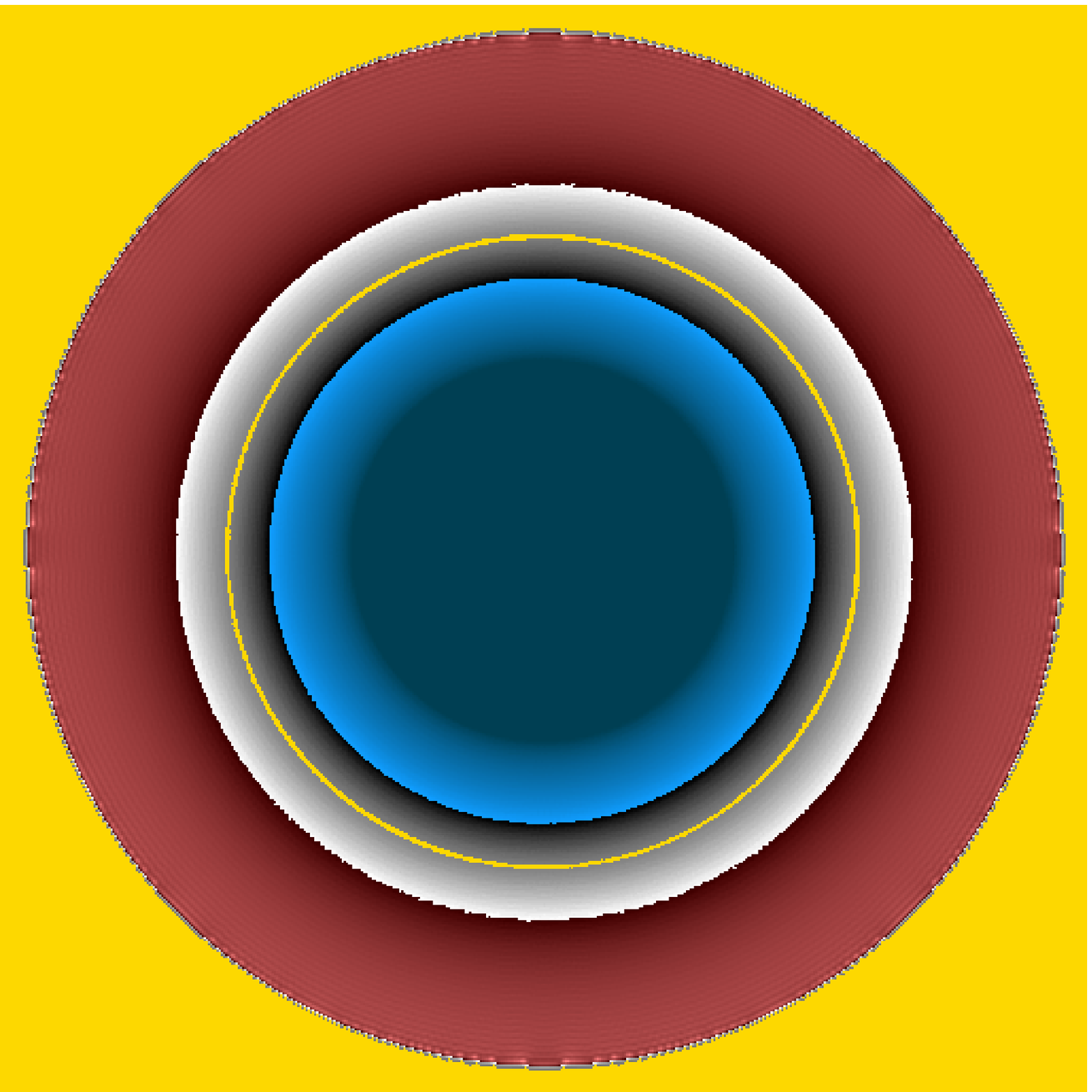}

 \includegraphics[scale=0.4,angle=270]{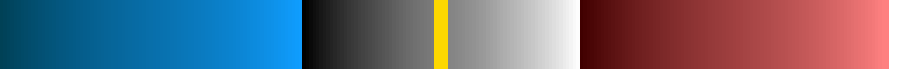}
\end{center}
\caption{\label{vort.per}\footnotesize{Current density at different instants in the circular domain. From top to bottom: regime I, regime II, regime III and regime IV; from left to right: $t=5$, $t=40$ and in the last column the time coresponds to $t=250$ for regime I and $t=1250$ for regimes II, III and IV.}\label{visucurr}}
\end{figure*}

Figures \ref{visuvort} and \ref{visucurr} show the vorticity and the current density field, respectively. For each of the cases I-IV, three typical time instants are visualized. These instants are $t=5$, showing the self-organization of the flow at early times, $t=40$, when nonlinear processes are dominating and $t=250$ (regime I) and $t=1250$ (regimes II, III and IV), corresponding to the final, viscously decaying state.

One flagrant feature of the visualizations is the local alignment of the magnetic and velocity field. Indeed in most regimes the vorticity and current density fields are rather similar. We also observe the coincidence of the maxima of $\omega$ and of $j$ which may have some effect on the stabilization of vorticity and current filaments. In case I an almost perfect axi-symmetrical state is achieved at $t=250$. Case II is the only case in which the formation of circular vortices is well pronounced, leading to a roll up of the current sheets. Apparently in the other regimes the Lorentz force suppresses the generation of circular vortices. Case III shows almost identical magnetic and velocity fields, as expected in this case of dynamic alignment, in which ${\bf u}$ and ${\bf B}$ are aligned (or anti-aligned) and in which kinetic and magnetic energies are in equipartition. Case IV is a typical example of the erratic regime: at the intermediate time, four dominant flow stuctures are observed, with both positive and negative cross-helicity. Locally the flow is close to an aligned or anti-aligned state, but globally the cross-helicity is weak because the different regions with opposite contributions cancel each other out.

\subsection{Final states}
\nointerlineskip

\begin{figure*}
\setlength{\unitlength}{0.8\textwidth}
\includegraphics[width=.3\unitlength]{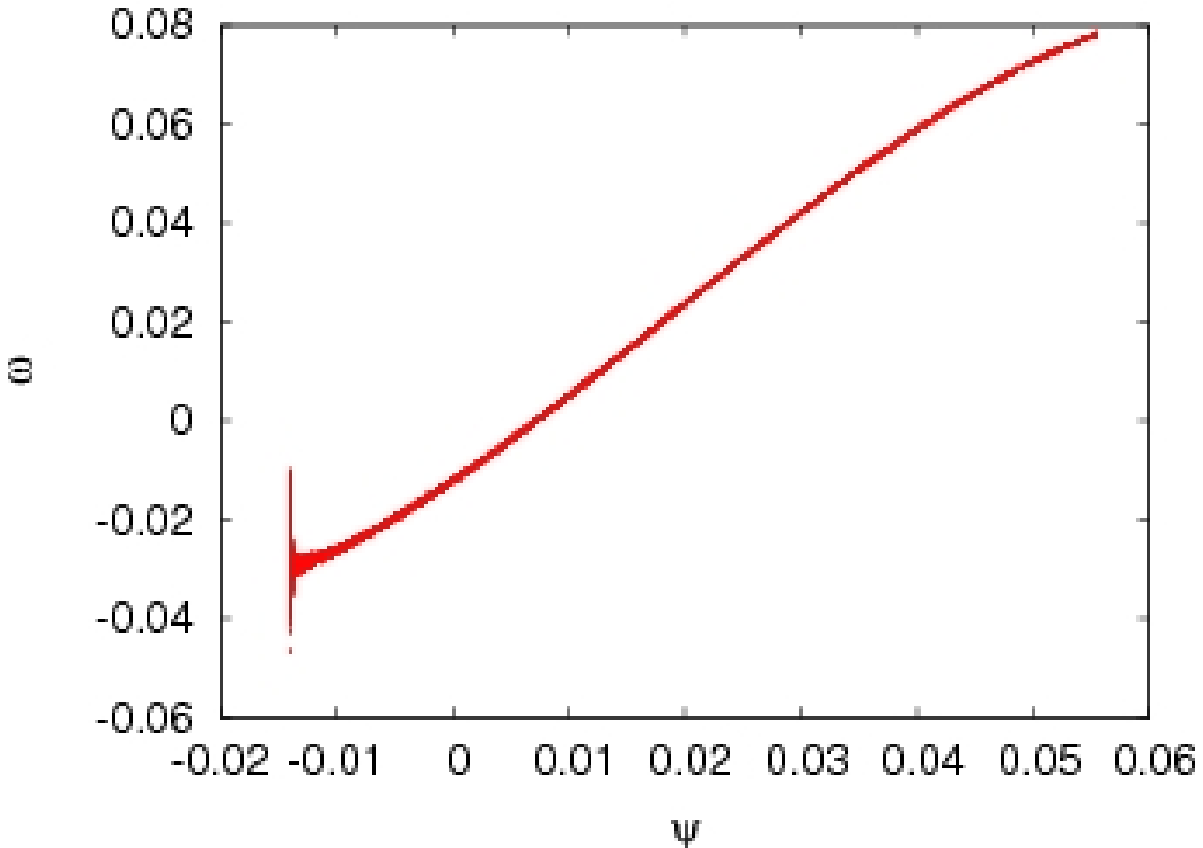}
\includegraphics[width=.3\unitlength]{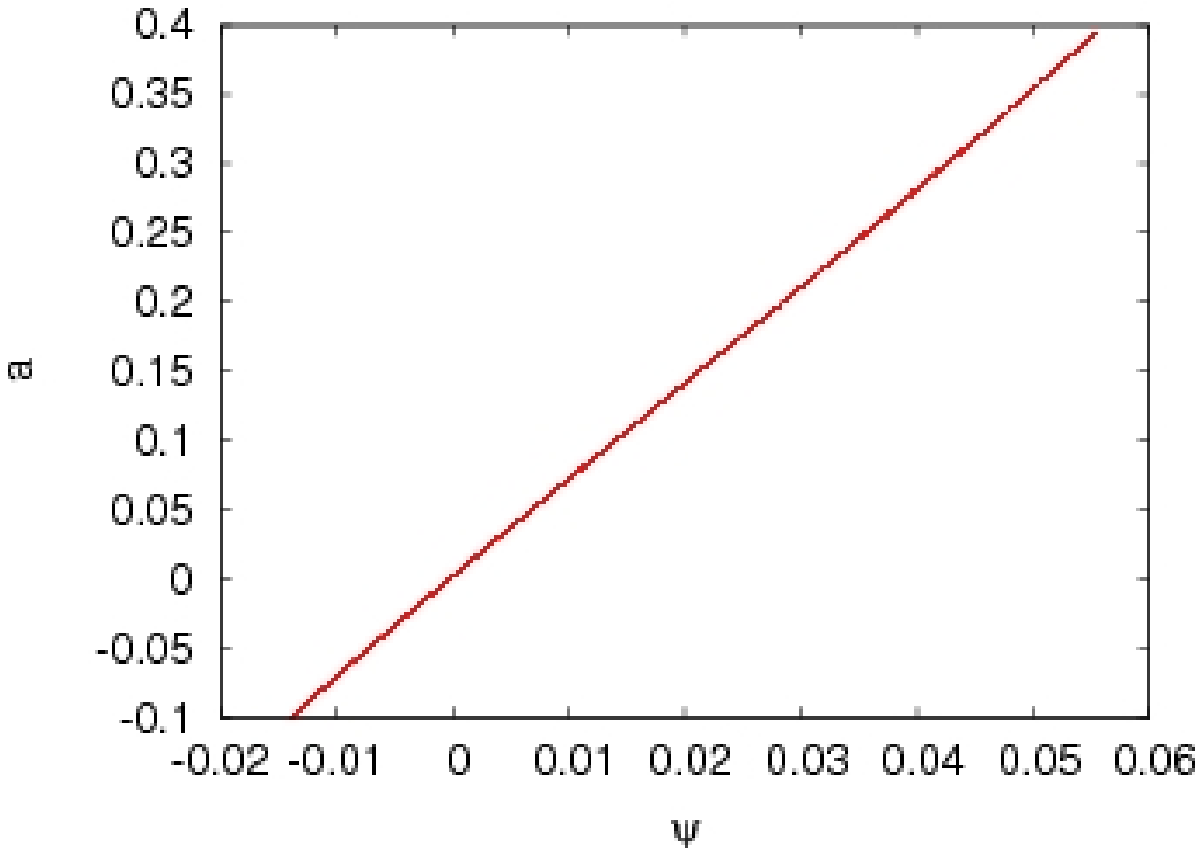}
\includegraphics[width=.3\unitlength]{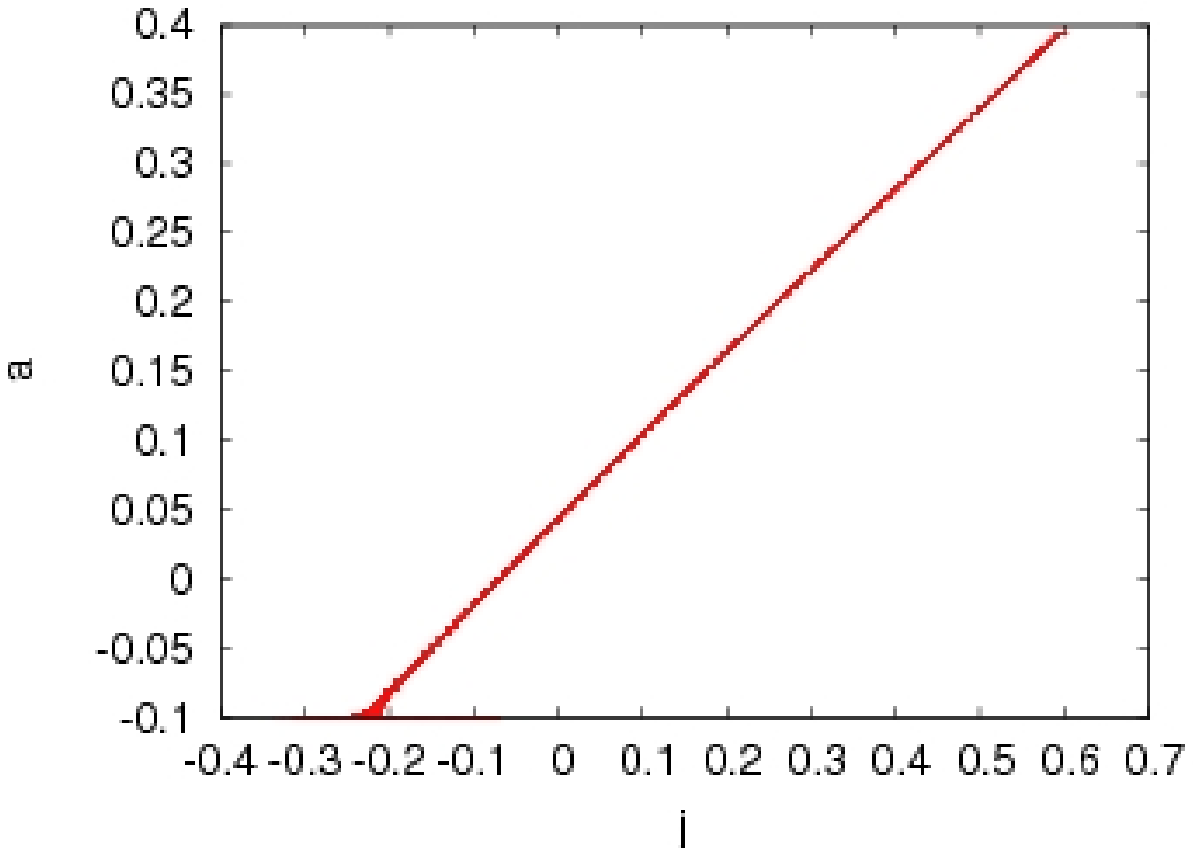}\\
\includegraphics[width=.3\unitlength]{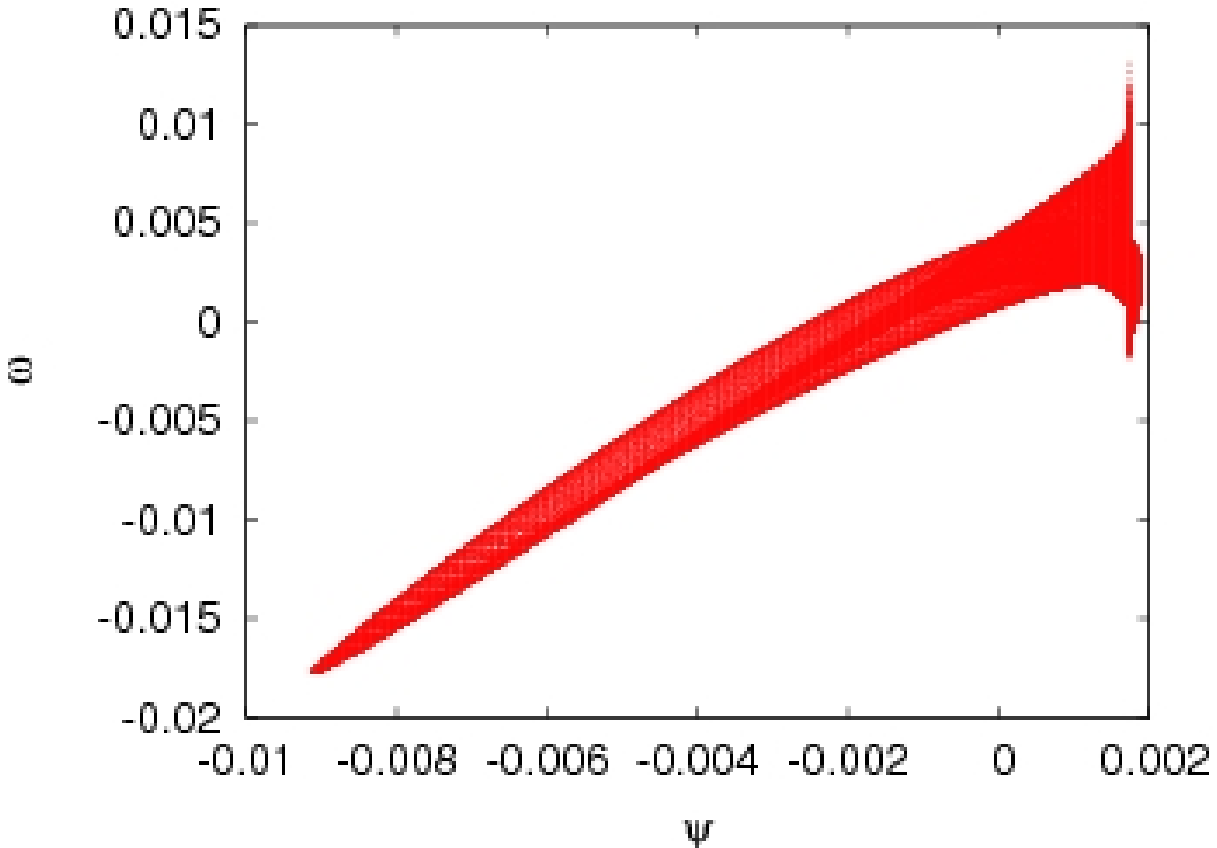}
\includegraphics[width=.3\unitlength]{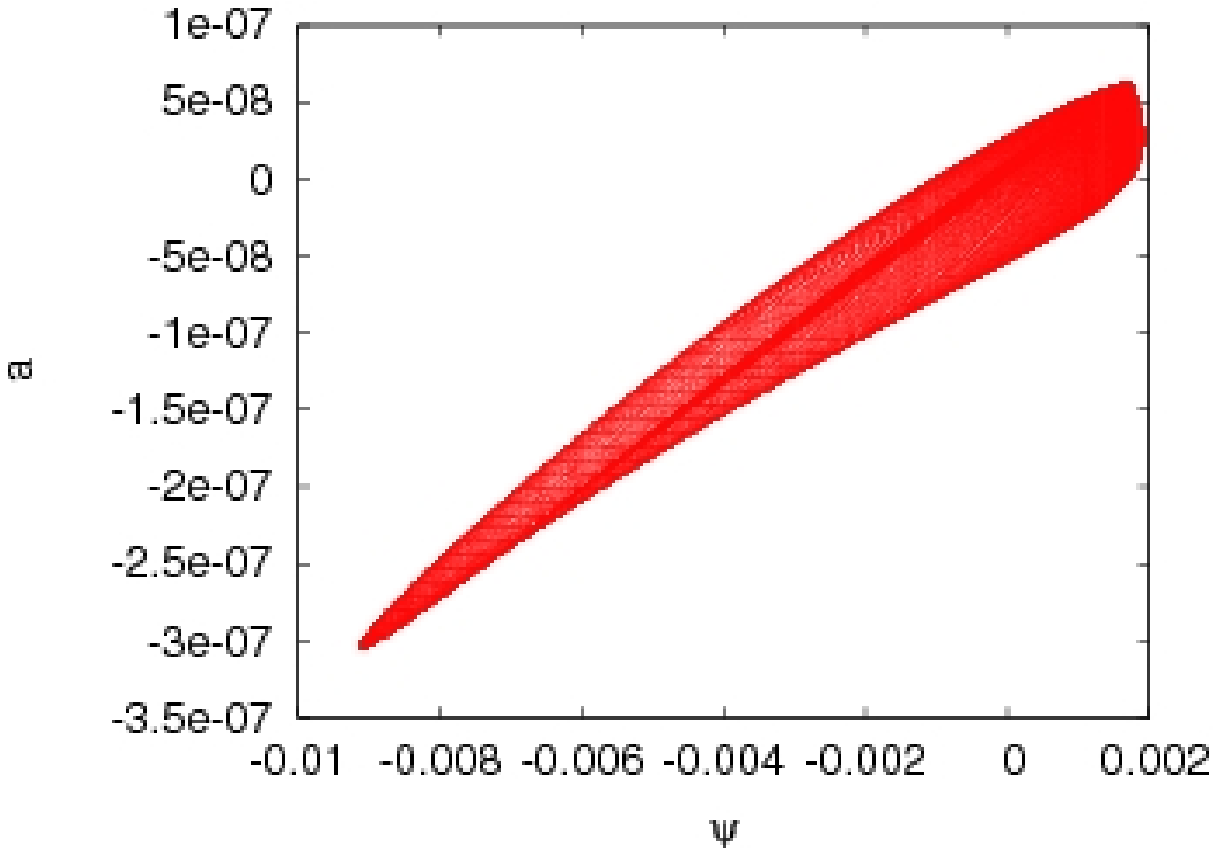}
\includegraphics[width=.3\unitlength]{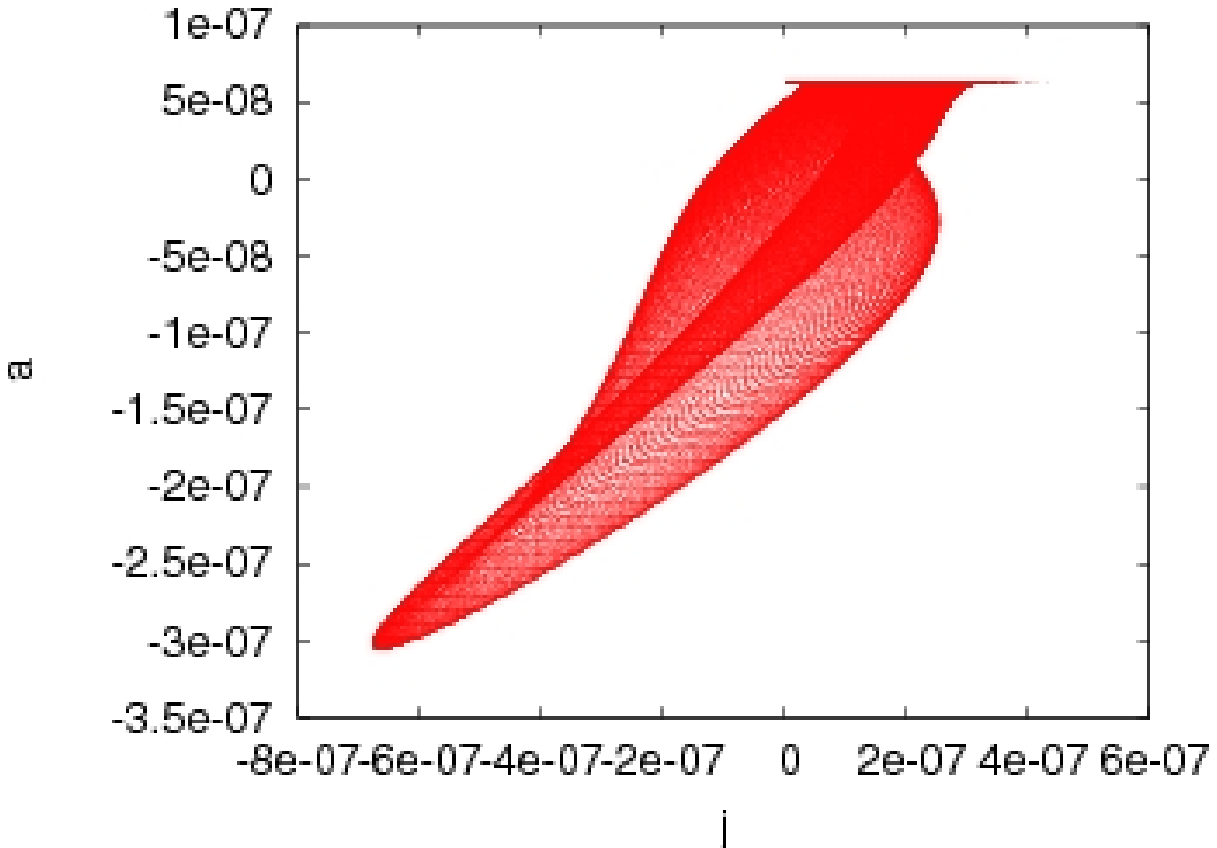}\\
\includegraphics[width=.3\unitlength]{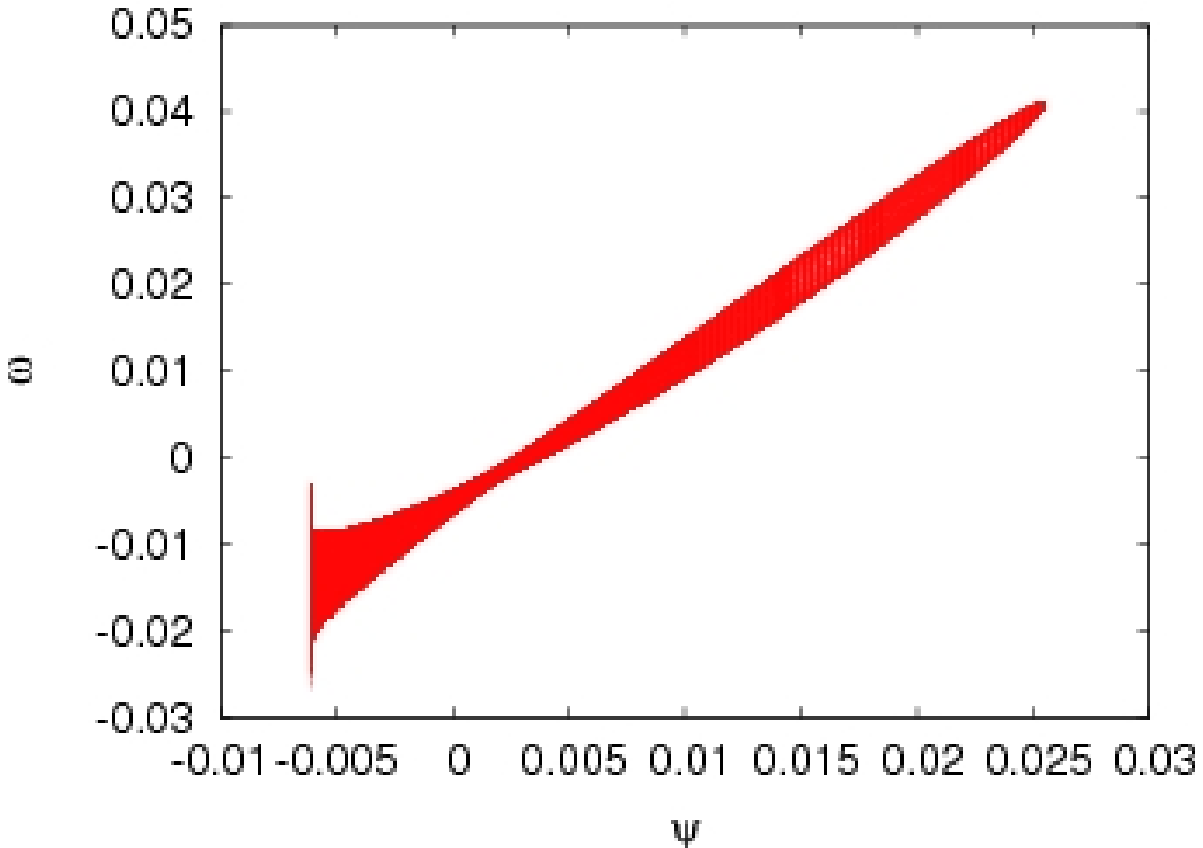}
\includegraphics[width=.3\unitlength]{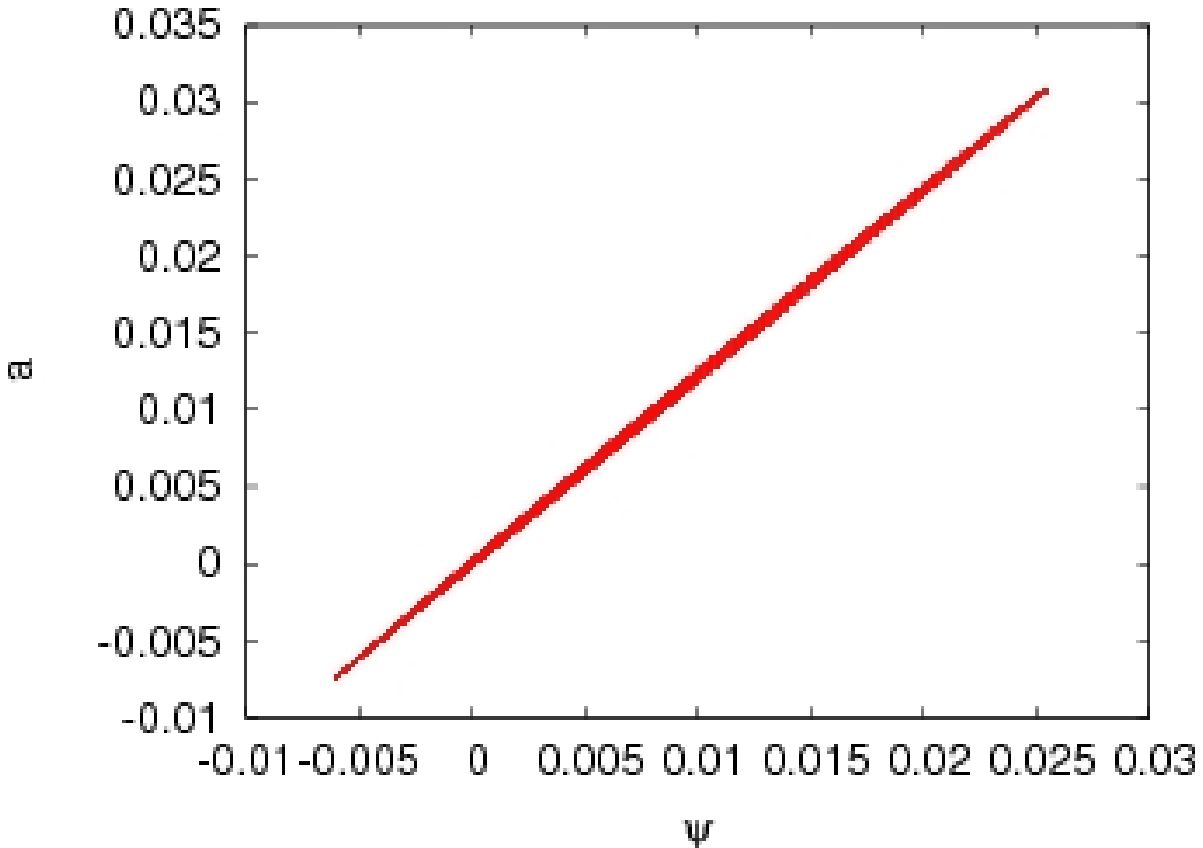}
\includegraphics[width=.3\unitlength]{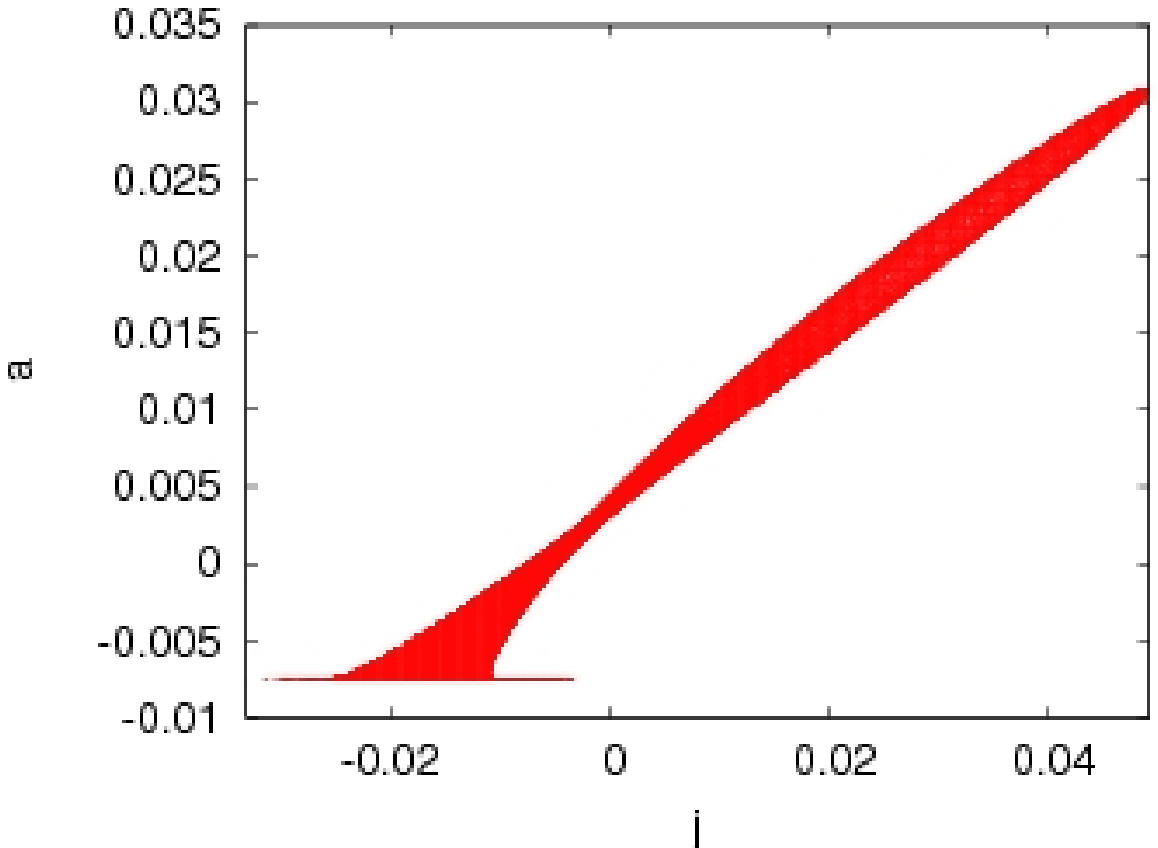}\\
\includegraphics[width=.3\unitlength]{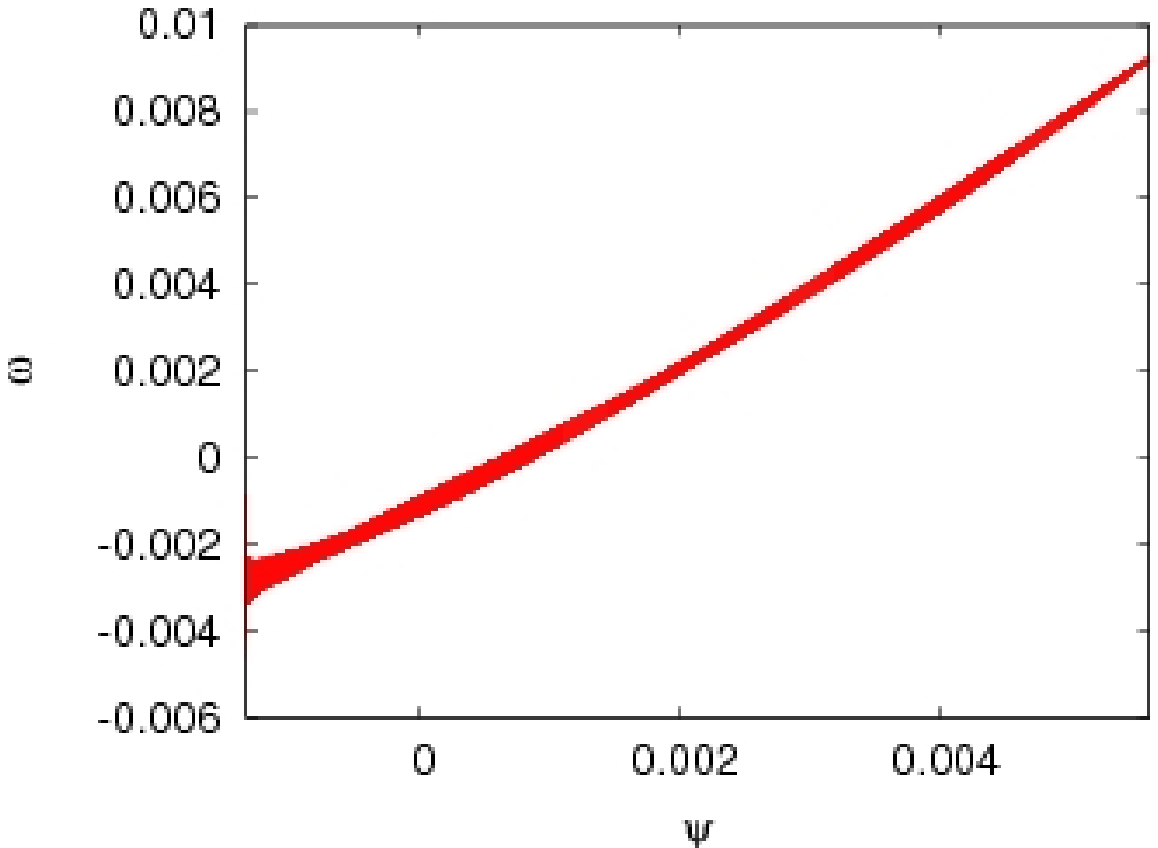}
\includegraphics[width=.3\unitlength]{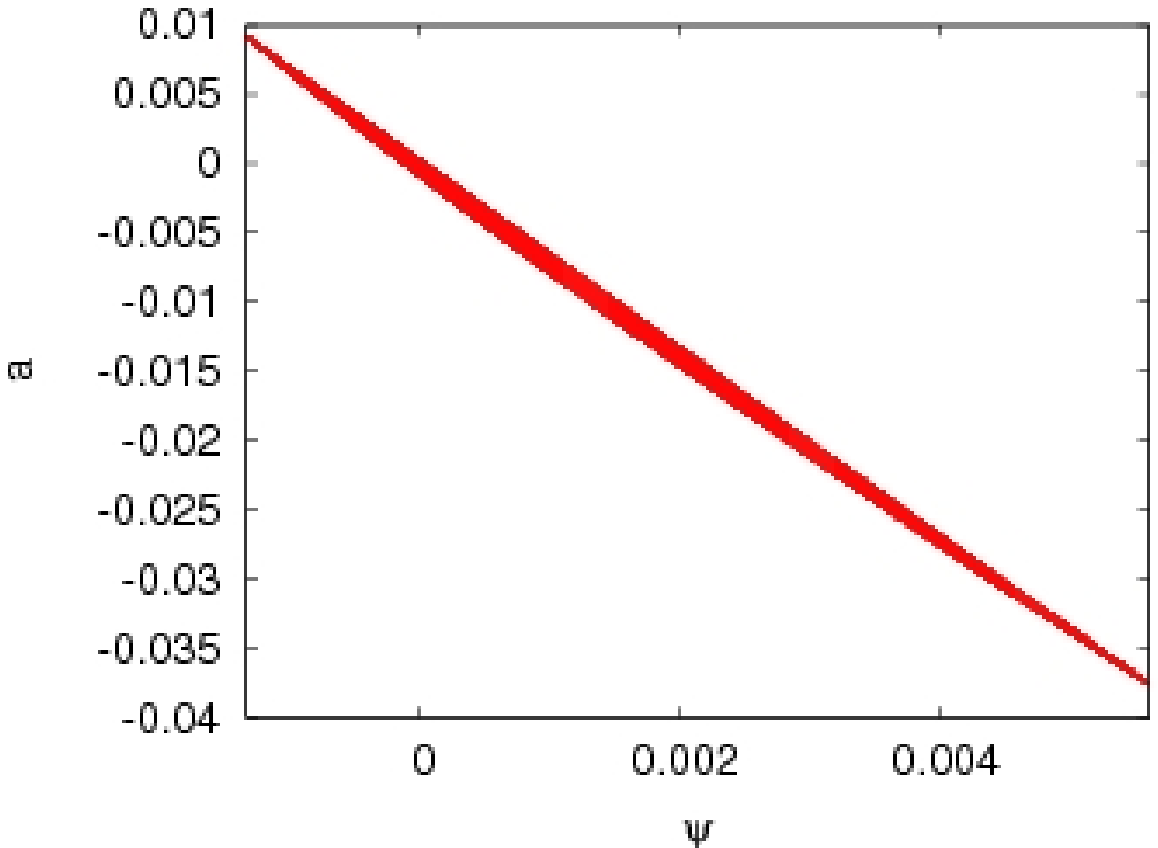}
\includegraphics[width=.3\unitlength]{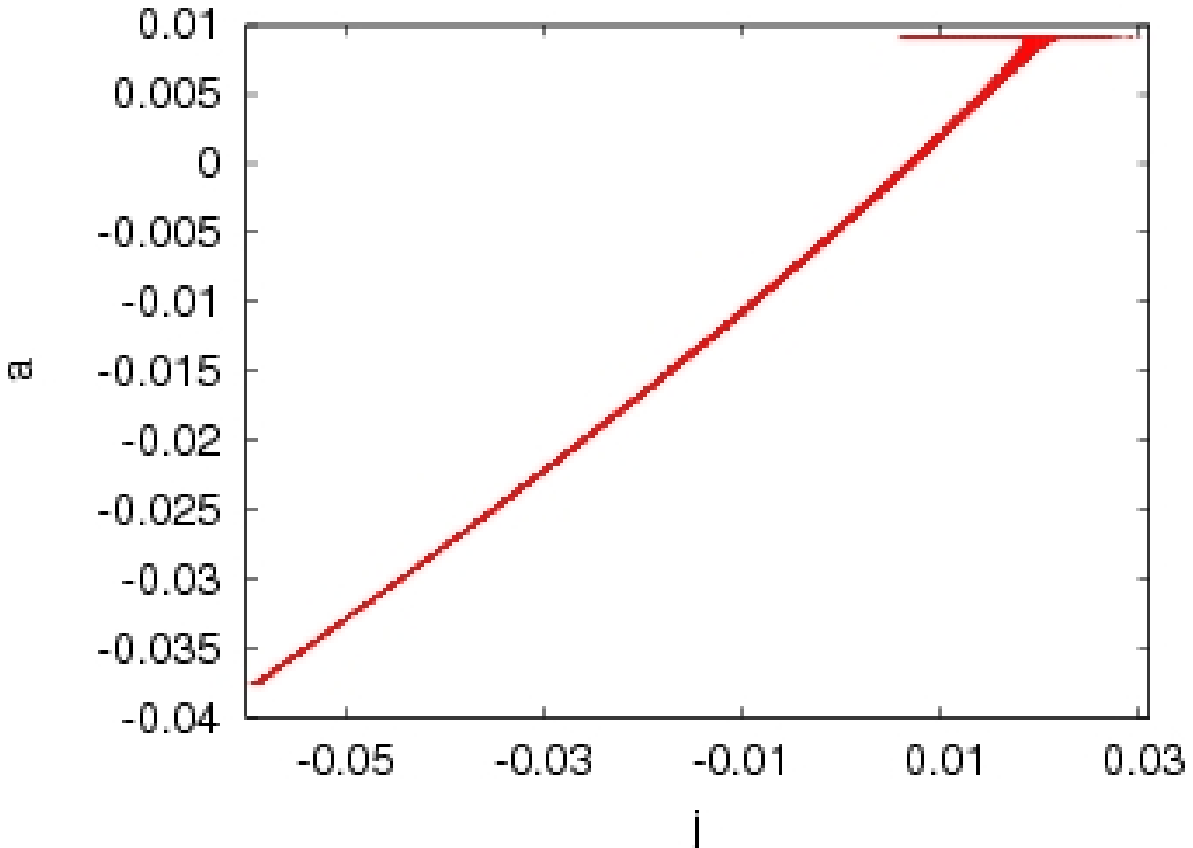}

\caption{Scatter plots of (from left to right) $\omega$ vs $\psi$, $a$ vs $\psi$ and $a$ vs $j$ for regimes (from top to bottom) I, II, III and IV at the latest time instant $t=250$ for regime I and $t=1250$ for regimes II, III and IV. \label{Scatter1}}
\end{figure*}

A supplementary information on the final states is given by scatter-plots.
It was shown by Joyce and Montgomery \cite{Joyce1973} that in hydrodynamic unbounded two-dimensional flows a long lasting final state is reached, depleted from nonlinearity. This state is characterized by a functional relation between the vorticity and the streamfunction of the form $\omega\sim \sinh(\psi)$. That a functional relation leads to a state, depleted from nonlinearity is easily shown from the equation for the vorticity:
\begin{eqnarray}
(\partial_t-\nu\Delta)\omega=[\omega,\psi],
\end{eqnarray}
with the Poisson bracket defined as $[a,b]=(\partial a/\partial x)(\partial b/\partial y)-(\partial a/\partial y)(\partial b/\partial x)$. A functional relation $\omega=F(\psi)$ leads to a vanishing Poisson bracket. If we consider now the equations for incompressible MHD:
\begin{eqnarray}
(\partial_t-\nu\Delta)\omega&=&[\omega,\psi]-[a,j]\\
(\partial_t-\eta\Delta)a&=&[a,\psi],
\end{eqnarray}
we see that two nonlinearities play a role: $[\omega,\psi]$ and $[a,j]$. The term $[a,\psi]$ can be considered as a pseudo-nonlinearity if $\psi$ is regarded as given. Although important theoretical progress has been made in the comprehension of final states \cite{Spineanu2003} no analytical nontrivial solution is presently known for the case of decaying MHD turbulence. It was however shown in Kinney \etal \cite{Kinney1995} that close to functional relations do exist in homogeneous two-dimensional MHD turbulence. In figure \ref{Scatter1} we show for the cases I-IV these scatter plots corresponding to the three nonlinearities.

In case I we see a well defined nonlinear functional relation $\omega(\psi)$. Clearly, we have a non trivial final state. The plot $a$ vs. $j$ shows a straight line, which corresponds to a vanishing Lorentz-force: the magnetic field does not interact with the velocity field at this final period of decay. The plot $a$ vs. $\psi$ also shows a clear functional relation. In case II, the scatter plots do not show such clear functional relations which is due to the fact that the flow is not yet sufficiently relaxed. The plot $\omega$ vs. $\psi$ is perhaps closest to a functional relation. In case III we see as expected a vanishing nonlinearity: for dynamic alignment it can be shown that nonlinearities vanish in the perfectly aligned case, when the equations are stated in Els\"asser variables (see for example \cite{Matthaeus1983}). In case IV it is expected that eventually the same behavior is observed as in case I. If the initial Reynolds number is initially too low this behavior will however not be observed. Preliminary computations were performed at lower resolution, which showed that non-trivial final states are only observed if the initial Reynolds number is sufficiently high. Otherwise linear relations are obtained for all different scatter plots.

\nointerlineskip
\section{Conclusion}
\nointerlineskip

We have investigated the influence of non-periodic boundary conditions on decaying two-dimensional magnetohydrodynamic turbulence. The use of a penalization method in combination with a classical Fourier pseudo-spectral method allows for efficient resolution of MHD flows in bounded domains. 

A main result is the observation of the robustness of the four different regimes discerned by Ting \etal \cite{Ting1986}. The same trends are found as in their pioneering work, depending on the initial values of the kinetic energy, magnetic energy, vector potential and cross-helicity. A detailed description was given of the relaxation-process which leads to the final states. In the case of a magnetically dominant, cross-helicity free case, a clear nontrivial functional relation was observed describing the magnetic and velocity fields. Functional relationships were also observed in regimes III and IV, while in regime II this functional relation was less clear.\\
Future work will address the influence of other types of boundary conditions for the magnetic field and also other geometries will be studied.\\

\noindent{\bf Acknowledgments:} We gratefully acknowledge Professor David Montgomery for his valuable remarks and we acknowledge financial support from the Agence Nationale de la Recherche, project ``M2TFP".

\end{document}